\documentclass{pnastwo1}

\usepackage[dvips]{graphicx}
\usepackage{amssymb,amsfonts,amsmath 
,color}
\usepackage{booktabs}
\usepackage{tabu}
\usepackage{nima,longtable1}

\begin{document}

\title{
Classical mechanics of economic networks}

\author{
Nima Dehmamy,~\affil{1}{Center for Polymer Studies, Boston University,
  Boston 02215, MA, USA} 
Sergey V. Buldyrev,~\affil{4}{
Department of Physics, Yeshiva University, New York, New York 10033, USA}
Shlomo Havlin,~\affil{3}{Bar-Ilan University, 52900 Ramat-Gan, Israel}
H. Eugene Stanley~\affil{1}{}
\and
Irena Vodenska,~\affil{2}{Administrative
Sciences Department, Metropolitan College, Boston University, Boston, MA
02215 USA,}\affil{1}{} 
}

\maketitle

\begin{article}

\begin{abstract}

Financial networks are dynamic. To assess their systemic importance
to the world-wide economic network and avert losses we need models that
take the time variations of the links and nodes into account. Using the methodology of classical mechanics and Laplacian determinism  we develop
a model that can predict the response of the financial network to a
shock. We also propose a way of measuring the systemic importance of the
banks, which we call BankRank. Using European Bank Authority 2011
stress test exposure data, we apply our model to the bipartite network
connecting the largest institutional debt holders of the troubled
European countries (Greece, Italy, Portugal, Spain, and Ireland). From simulating  our model we can determine whether a
network is in a ``stable'' state in which shocks do not cause major
losses, or a ``unstable'' state in which devastating damages
occur. 
Fitting the parameters of the model, which play the role of physical coupling constants, to Eurozone crisis data shows that before
the Eurozone crisis the system was mostly in a ``stable''  regime, and that
during the crisis it transitioned into an ``unstable'' regime. The
numerical solutions produced by our model match closely the actual
time-line of events of the crisis.  We also find that, while the largest
holders are usually more important, in the unstable regime smaller holders also
exhibit systemic importance. 
\outNim{
In addition, we observe that asset
diversification has no clear correlation with our BankRank. 
Thus diversification neither reduces systemic contagion, nor necessarily
provides routes for contagion.
}
Our model also proves useful for determining the vulnerability of banks and assets to
shocks. 
This suggests that our model may be
a useful tool 
 for simulating the response dynamics of shared portfolio networks.

\end{abstract}


\subsubsection*{Significance}

We propose a simple yet powerful deterministic model for a fully dynamical bipartite
network of banks and assets and apply it to the Eurozone sovereign debt crisis.  
The results closely match real-world events
(e.g., the high risk of Greek sovereign bonds and the failure of Greek
banks).  The model can be used to conduct ``systemic stress tests'' to
determine the vulnerability of banks and assets in time-dependent
networks.  It also provides a simple way of assessing the
stability of a system by using the ratio of the log returns of sovereign
bonds and the stocks of major holders. We also propose a ``systemic
importance'' ranking, BankRank, for these dynamic bipartite networks.

\bigskip

\dropcap{R}ecent financial crises have motivated the scientific
community to seek new interdisciplinary approaches to modeling the
dynamics of global economic systems. Many of the existing economic
models assume a mean-field approach, and although they do include noise
and fluctuations, the detailed structure of the economic network is
generally not taken into account. Over the past decade there has been
heightened interest in analyzing the ``pathways of financial
contagion.''  The seminal papers were by Allen and Gale
\cite{Contagion,Gale2} and these were following by many other studies
\cite{Furfine,Wells,Upper,Elsinger,Nier,Cifuentes}. Economists have
recently become aware that econometrics has traditionally paid
insufficient attention to two factors: (i) the structure of economic
networks and (ii) their dynamics.  Studies indicate that a more thorough
approach to the examination of economic systems must necessarily take
network structure into consideration
\cite{Haldane,Moreno,Johnson,Schweitzer,Watts,Sergey,Dror1,Dror2}.

One example of this approach is the work of Battiston et
al. \cite{Battiston}. They study the 2008 banking crisis and use network
analysis to develop a measure of bank importance. By defining a dynamic
centrality measurement called DebtRank that measures interbank lending
relationships and their importance in propagating network distress, they
show that the banks that must be rescued if a crash is to be avoided
(those that are ``too big too fail'') are the ones that are more
``central'' in terms of their DebtRank.

Another recent event that has motivated and provided the focus for our
study reported here is the 2011 European Sovereign Debt Crisis. It began
in 2010 when the yield on the Greek sovereign debt started to diverge
from the sovereign debt yield of other European countries, and this led
to a Greek government bailout \cite{Lane2012}. The nature of the
sovereign debt crisis and resulting network behavior that we analyze
here differs somewhat from that of the US banking crisis. Here we focus
on the funds that several Eurozone countries---Greece, Italy, Ireland,
Portugal, and Spain (GIIPS)---had borrowed from the banking system
through the issuing of bonds. When these governments faced fiscal
difficulties, the banks holding their sovereign debt faced a dilemma:
should they divest some of their holdings at reduced values or should
they wait out the crisis.  The bank/sovereign-debt network that we
analyze in this study is a bilayer network. Although DebtRank has also
been used to study bipartite networks, e.g., to describe the lending
relationships between banks and firms in Japan \cite{Hideaki}, it does
not take into account that link weights exhibit a dynamic behavior.

Huang et al. \cite{Huang} and Caccioli et al. \cite{Caccioli} analyzed a
similar problem, that of cascading failure in a bipartite network of
banks vs assets in which risk propagates among banks through overlapping
portfolios (see also Ref.~\cite{Tsatskis}). 
\outNim{ 
The interesting conclusion
of these analyses is that, contrary to the common economic assumption,
portfolio diversification does not always reduce risk and can in fact
provide new pathways for risk propagation. The results produced by these
analyses using network simulation run counter to the common economic
wisdom regarding diversification (see also Ref.~\cite{Chollete}).
} 
Although network connections in real-world financial systems, e.g.,
interbank lending networks or stock markets, are dynamic, neither of the
above models \cite{Huang,Caccioli} take this into account.  Other models
by Ha\l aj and Kok \cite{Kok}, which use simulated networks similar to
real systems, or by Battiston et al. \cite{Liaison} allow the nodes to
be dynamic but not the links (see, however, Ref.~\cite{Kok2}, in which
dynamic behavior occurs when a financial network attempts to optimize
``risk adjusted'' assets \cite{Kok3}).  Our approach differs from both
of these because by introducing only two parameters which play the role of coupling constants in physics we can enable all
network variables to be dynamic.  Our model is related to Caccioli et
al. \cite{Caccioli} and Battiston et al. \cite{Battiston} but differs in
that we allow both nodes and links to be dynamic.

We use a time-slice of the GIIPS sovereign debt holders network from the
end of 2011 to focus on a simplified version of the network structure
and use it to set the initial conditions for our model.\footnote{To
  assess the robustness of our model, we use what we learn about the
  GIIPS network structure and apply it to simulations of other networks
  of varying sizes and distributions. These results are available in the
  supporting materials.}

We start by proposing, solely on phenomenological grounds, a set of
dynamical equations.  Based on our analysis we observe that:

\begin{enumerate}

\item When we model how a system responds to an individual bank
  experiencing a shock, our analysis is in accordance with real-world
  results, e.g., in our simulations Greek debt is clearly the
  most vulnerable.

\item The dynamics arising from our model produces different end
  states for the system depending on the values of the parameters.

\end{enumerate}

\noindent
In order to determine which banks play a systemically dominant role in
this bipartite network, we adjust the equity of each bank until it goes
bankrupt and then quantify the impact (the BankRank) of the bank's
failure on the system.  We simulate the dynamics for different parameter
values and observe that the system exhibits at least two distinctive
phases, one in which a new equilibrium is reached without much damage
and one in which the monetary damage is quite significant, even
devastating.

\section{The GIIPS problem}

Governments borrow money by issuing sovereign (national) bonds that trade in a bond market (which is similar to a stock market\footnote{The entity that issues a bond (e.g., the  government in case of sovereign bond) promises to pay interest. Governments also promise to return the face value of the  loan at the ``maturity'' date. Bonds, unlike stocks, have maturities
  and interest payments.  A detailed description of some of these bond
  characteristics can be found in Ref.~\cite{Battiston2}.  As is the
  case with stocks, the value of these sovereign bonds increases when
  countries are doing well, and supply and demand ultimately determine
  the value of the bonds. If, however, the country becomes troubled and
  the market perceives that the government will not be able to pay back
  the debt, the price of the bond can crash, which was the case of
  Greece.}).

\begin{figure}
\centerline{\includegraphics[width=1.7in]{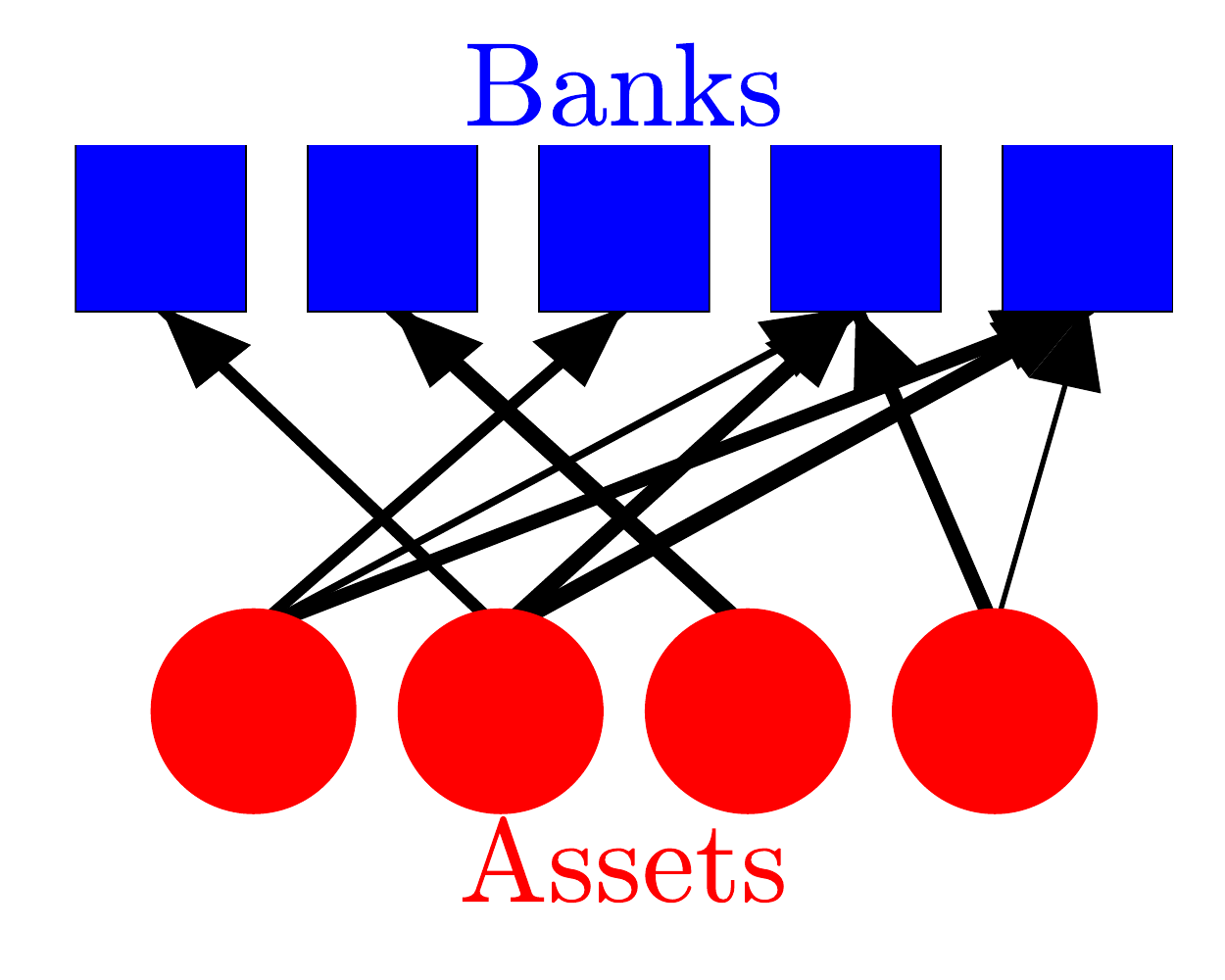}\includegraphics[width=1.9in]
{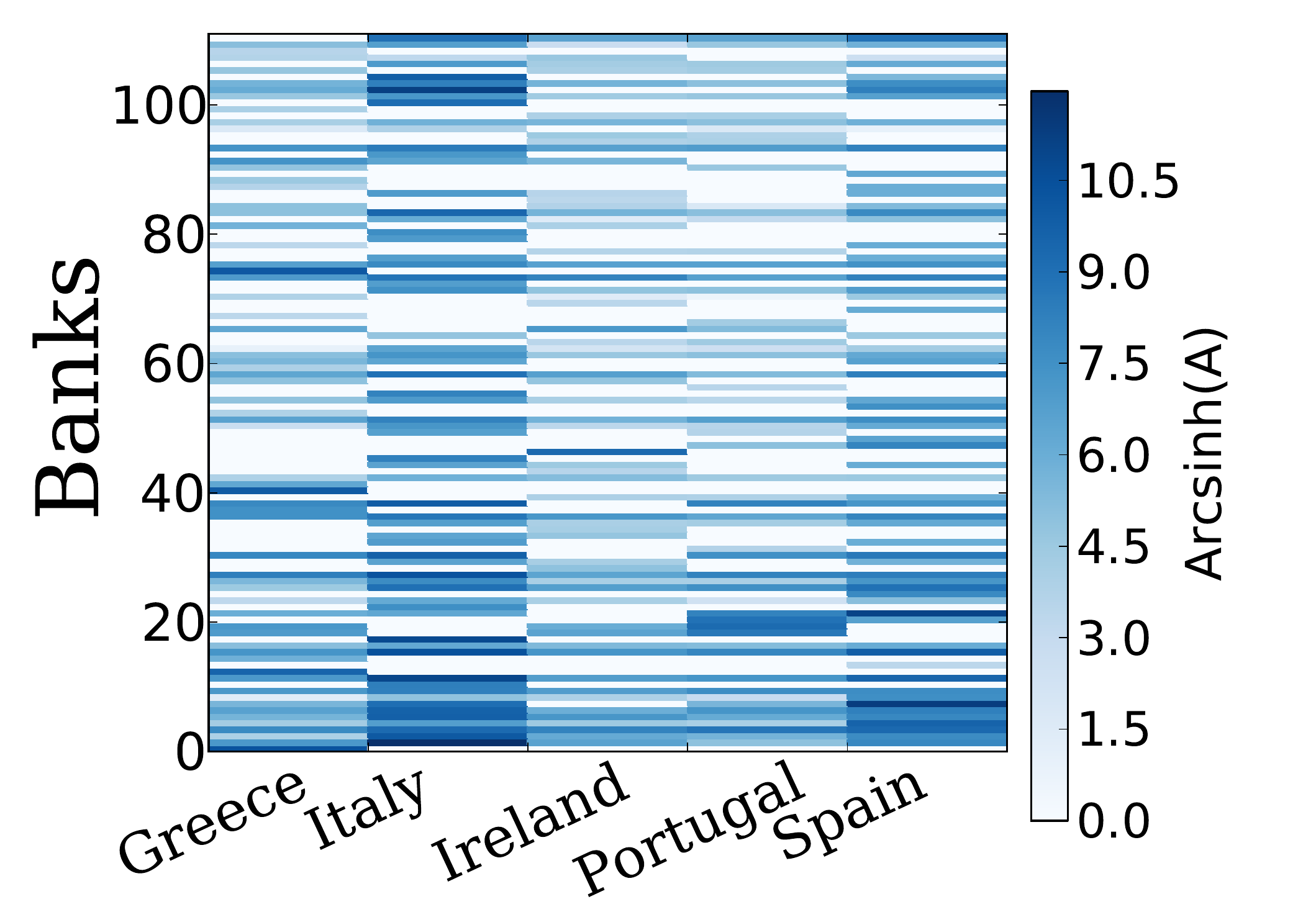}}
\caption{(Left) A sketch of the network of banks vs assets. It is a
  directed, weighted bipartite graph. The thicknesses represent holding
  weights. Motion along the edges from banks to assets is described with
  the wighted adjacency matrix $A$ and the opposite direction, assets to
  banks, is described with $A^T$. \label{fig:graph} (Right) $\sinh^{-1}
  (A)$ with $A$ being the weighted adjacency matrix of the GIIPS
  holdings, (weighted by amount of banks' holdings in GIIPS sovereign
  debt expressed in units of millions of Euros.  The vertical axis
  denotes different banks (121 of them) and they are ordered in terms of
  their total exposures to GIIPS debt (higher exposure is at the bottom
  of the plot) Because holdings differ by orders of magnitude we have
  plotted $\sinh^{-1} A$ here.
\label{fig:adj} }
\end{figure}

Our GIIPS data are from the 137 banks, investment funds, and insurance
companies that were the top holders in the GIIPS sovereign bond-holder
network in 2011. (Hereafter we will use ``banks'' to refer to all these
financial institutions.) Table~\ref{tab:debt} shows the percentages of
the sovereign bonds issued by each GIIPS country owned by these
banks. Since our model requires knowing the equity of each bank, we
reduce our dataset to the 121 banks whose equity value was
obtainable. By the end of 2011, two important Greek banks---the National
Bank of Greece and Piraeus Bank---had negative equities. Because our
model only considers banks that can execute trades based on positive
capital, we also had to eliminate these two banks from our analysis.
Figure~\ref{fig:adj} shows the weighted adjacency matrix of this
network.\footnote{The intensity of the color is proportional to
  Arcsinh($A$) for better visibility. For large $A_{i\mu}$,
  arcsinh$(A)\approx log(2 A)$.}

\begin{table}
\caption{Total amount of exposure of the banks in our data set to the
  sovereign debt of the GIIPS countries\label{tab:debt}}
\centering
\small
\begin{tabular}{@{\vrule height 8.5pt depth4pt  width0pt}lccccc}\hline
& Greece & Italy & Portugal & Spain & Ireland\\ \hline
Total (bnEu) & 64.85 &330.38
& 30.63&151.15 &18.41\\ \hline
\% in Banks & 23.67 &20.13 &23.81 & 21.81 &20.55\\ \hline
\end{tabular}
\end{table}

When a country defaults on sovereign debt (or stops paying interest as
it comes due) the consequences are usually grave.  To prevent cascading
sovereign defaults, the European Union, the European Central Bank, and
the International Monetary Fund jointly established financial programs
to provide funding to troubled European countries. Funding was
conditional on implementing austerity measures and stabilizing the
financial system in order to promote growth and increase productivity.
We use our sovereign debt data as the initial condition for a model of
cascading distress propagating through a bipartite bank network in which
banks only affect each other through shared portfolios. In order to
develop a framework for analyzing these problems that goes beyond simply
determining how distress propagates through the links, we construct a
model in which dynamic change affects both the weights of links and the
attributes of nodes. Figure~\ref{fig:adj} shows the weighted adjacency
matrix of this network in log format.

\section{Model}

The system that we study is a bipartite network as shown in
Fig.~\ref{fig:graph}.  On one side we have the GIIPS sovereign bonds,
which we call ``assets,'' and on the other we have the ``banks'' that
own the GIIPS bonds. The nodes on the ``asset'' side are labeled using
Greek indices $\mu,\nu...$.  To each asset $\mu$ we assign a ``price,''
$p_\mu(t)$ at time $t$. The ``bank'' nodes are labeled using Roman
indices $i,j...$. Each bank node has an ``equity'' $E_i(t)$, a time $t$,
and an initial value of asset $\mu$. Each bank in the network can have
differing amounts of holdings in each of the asset types. The amount of
asset $\mu$ that bank $i$ holds is denoted by $A_{i\mu}(t)$, which is
essentially an entry of the weighted adjacency matrix $A$ of the
network.  In our model we begin with a set of phenomenological equations
describing how each of the variables $E_i(t), A_{i\mu}(t)$, and
$p_\mu(t)$ evolve over time. A key feature of our model is that the
weights of links $A_{i\mu}$ are time-dependent, and this introduces
dynamics into our network.

\subsection{Assumptions, simplifications and the GIIPS system\label{sec:GIIPS}}

The key assumptions that differentiate our model from other banking
system or dynamic network models are:

\begin{enumerate}

\item The banks do not {\it exclusively\/} trade with each other. They
  may trade with an external entity, which may be the European Central
  Bank (ECB) or other, smaller investors.
  \footnote{This is appropriate in the case of GIIPS sovereign debt
    because, in addition to the ECB (which buys some of the bonds if
    there is a need to stabilize the system), a large number of
    investors hold GIIPS sovereign debt. This is important to keep in
    mind because in most problems associated with banking or financial
    networks agents are assumed to be trading with each another.}

\item When there is no change in equity, price, or bond holdings,
  nothing happens and there is no intrinsic dynamic activity in our
  financial network. \label{en:static}

\outNim{
\item The attitude (panic level) is the same for all banks and does not
  change during the period considered (i.e., during a short-term
  crisis).

\item The response of the market (the ``inverse market depth'' factor) is
  also constant during this period.
 } 
\item The model describes the short time response of the system
and disregards slow, long-term driving forces of the market.
\item We assume the agents in the system will copy each other’s
actions, producing the so-called ``herding effect.'' This is
why we assume the ``coupling constants'' (the free parameters) 
are the same for all agents.

\end{enumerate}


\section{Notations and Definitions}

We denote by $A$ the weighted adjacency matrix, the components of which
$A_{i\mu}$ are the amount of exposure of bank $i$ to asset $\mu$.  The
equity $E_i$ of a bank is defined as
\[E_i= \sum_\mu A_{i\mu}p_\mu +C_i-L_i.\]
Here $p_\mu$ is the ``price ratio'' of asset $\mu$ at a given time to
its price at $t=0$, $C_i$ is the bank's cash, and $L_i$ is bank's
liability. These parameters evolve in time. Bank $i$ will fail if its
equity goes to zero,
\[if: \quad E_i=0 \quad \to\quad \mbox{Bank $i$ fails.}\]
We assume that the liabilities are independent of the part of the market
we are considering and are constant. For convenience we define
\[c_i\equiv C_i-L_i. \]

Two other dependent variables that we use are the ``bank asset value''
$V_i\equiv \sum_\mu A_{i\mu}p_\mu$ and the total GIIPS sovereign bonds
on the market $A_\mu\equiv \sum_i A_{i\mu}$.
%
%
In our model we assume that $\alpha$ and $\beta$ are
constant. Everything else is time-dependent.

\subsection{The time evolution of GIIPS holdings and their price}

For changes in equity we have
\[\delta E_i=\sum_\mu\pa{(\delta A_{i\mu})p_\mu +A_{i\mu} \delta p_{\mu}} +\delta c_i.\]
Here we assume that the cash minus liability changes according to the
amount of money earned through the sale of GIIPS holdings,
\[
\delta c_i= -\sum_\mu (\delta A_{i\mu}) p_\mu+\delta S_i(t),
\]
where the minus sign indicates that a sale means $\delta A_{i\mu}<0$ and
this should add positive cash to the equity of bank $i$. $\delta S_i(t)$ is the
cash made from transactions outside of the network of $A_{i\mu}$. The
first term in $\delta c_i$ cancels one term in $\delta E_i$ and we get
(all at time $t$)
\[\delta E_i= \sum_\mu A_{i\mu} \delta p_\mu+\delta S_i(t). \]
In the secondary market for the bonds (where issued bonds are traded in
a manner similar to stocks) the prices are primarily determined by
supply and demand.  We use a simple model for the pricing that should
hold as a first-order approximation. We assume the price changes to be
\[\delta p_\mu(t+\tau_A)=\alpha {\delta A_\mu(t)\over A_\mu (t)}p_\mu (t),\]
%
Here the coupling constant $\alpha$ is essentially the  ``inverse of the market depth,'' i.e. the fraction of sales ($\delta A/A$ required to reduce the price by one unit ($\delta p /p$) is equal to $1/\alpha$.  We are assuming that the market is ``liquid'' meaning that any amount of assets can be sold or bought without a problem. 
We have defined $\delta p_\mu(t)\equiv p_\mu(t)-p_\mu(t-\delta t)$ is the change
in price from the previous step, $\delta
A_\mu(t)=A_\mu(t)-A_\mu(t-\delta t)$ the net trading (number of
purchases minus sales) of asset $\mu$, and $\tau_A$ the ``response time
of the market.''  
 We choose the same ``inverse market depth'' for all
GIIPS holdings $\mu$, assuming that they belong to the same class of
assets.  We then define how the GIIPS holdings are sold or bought, i.e.,
we define $\delta A_{i\mu}$. 
We also include a ``panic factor'' $\beta$ that indicates how abruptly
distress propagates when a loss is incurred, and a ``market
sensitivity'' factor $\alpha$ that indicates how quickly the price of an
asset drops when part of it is sold. These variables are summarized in
Table~\ref{tab:definitions}.

\begin{table}
\caption{Notation\label{tab:definitions}}
\centering
\begin{tabular}{@{\vrule height 10.5pt depth4pt  width0pt}cc}
\hline
symbol & denotes \\
\hline
$A_{i\mu}(t)$ & Holdings of bank $i$ in asset $\mu$ at time $t$\\
$p_\mu (t)$ & Normalized price of asset $\mu$ at time $t$ ($p_\mu(0)=1$)\\
$E_i(t)$ & Equity of bank $i$ at time $t$. \\
$\beta$ & Banks\rq{}  ``Panic'' factor.\\
$\alpha$ & ``Inverse market depth'' factor of price to a sale.\\
\hline
\end{tabular}
\end{table}

We assume that if a bank's equity shrinks
it will start selling GIIPS holdings in order to continue meeting its
liability obligations, and that if a bank's equity shrinks because of
asset value deterioration it will sell a fraction of its entire
portfolio to ensure meeting those obligations. The amount of GIIPS
holdings sold will depend on how panicked the bank is, i.e., on the
value of ``panic factor'' $\beta$. A bank thus determines what fraction
of its equity has been lost in the previous step and sells according to
\[\delta A_{i\mu}(t+\tau_B)=\beta {\delta E_i(t)\over E_i(t)} A_{i\mu}(t),\]
where $\tau_B$ is the ``response time of the banks.'' Here we assume
that banks purchase using the same protocol as when selling and sell the same fraction of all their
GIIPS assets.  
\outNim{ 
In summary, we propose the following equations for the
dynamics.\footnote{Without a time lag, these equations would be
  primarily constraint equations relating the first-order time
  derivatives of $E,p,A$ to each other. Note however that in simulating
  this dynamic system the order in which we update the variables matters
  because most of the nontrivial dynamic behavior follows from this time
  lag between updates.} We write
\begin{align}
\delta A_{i\mu}(t+\tau_B)&=\beta {\delta E_i(t)\over E_i(t)} A_{i\mu}(t)\label{eq:dA}\\
\delta p_\mu(t+\tau_A)&=\alpha {\delta A_\mu(t)\over A_\mu (t)}p_\mu (t)\label{eq:dp} \\
\delta E_i(t)&= \sum_\mu A_{i\mu}(t) \delta p_\mu(t)+f_i(t). \label{eq:dE}
\end{align}
} 
The above equations can be converted to differential equations by simply
replacing $\delta F \to dF/dt$. If we assume that the time lags are
small, we can expand the equations with $\tau_A,\tau_B$ to first-order
and get
\[{dF(t+\tau)\over dt}\approx {d\over dt}\pa{F(t)+\tau {dF\over dt}}\]
For brevity, we define $\ro_t\equiv {d\over dt}$. The three equations become:
\begin{align}
\pa{\tau_B\ro_t^2 +\ro_t }A_{i\mu}(t)&=\beta {\ro_t E_i(t)\over E_i(t)} A_{i\mu}(t)\label{eq:ddA}\\
\pa{\tau_A\ro_t^2 +\ro_t } p_\mu(t)&=\alpha {\ro_t A_\mu(t)\over A_\mu (t)}p_\mu (t)\label{eq:ddp} \\
\ro_t E_i(t)&= \sum_\mu A_{i\mu}(t) \ro_t p_\mu(t)+f_i(t). \label{eq:ddE}
\end{align}
where $f_i=dS_i/dt$ has the meaning of external force. 
where $\tau_B$ is the time-scale in which Banks respond to the change,
and $\tau_A$ is the time-scale of market's response.\footnote{Without a
  time lag, these equations would be primarily constraint equations
  relating the first-order time derivatives of $E,p,A$ to each
  other. Note however that in simulating this dynamic system the order
  in which we update the variables matters because most of the
  nontrivial dynamic behavior follows from this time lag between
  updates.}


In our simulations we use these differential equations and choose
$\tau_A=\tau_B=1$. One of them can always be chosen as a time unit and
set to one, but setting them equal is an assumption and may not be true
in reality. 
Our analysis showed that the choice of $\tau_{A,B}$ does not
affect the stability of the system and that the stability only depends
on $\alpha$, $\beta$ and the shock. The $f_i(t)$, which are changes in
the equity from what banks own outside of this network, can be thought
of as external noise or driving force. We use $f_i(t)$ to shock the
banks and make them go bankrupt. We shock a single bank, say bank $j$,
at a time by reducing its equity 10\% by putting\footnote{Note that the
  magnitude of the shock only rescales time, according to
  Eq.~\eqref{eq:ddE} because $f_i\to \lambda f_i$ is the same as
  $\ro_t\to \lambda^{-1}\ro_t$ and thus $\tau_{A,B} \to \lambda
  \tau_{A,B}$} $f_i(t)=s E_j\delta_{ij}
\delta(t)$.\footnote{$\delta_{ij}$ is the kronecker delta, or the
  identity matrix elements, and $\delta(t)$ is the Dirac distribution or
  impulse function.} Starting with $\ro_t p_\mu(-\eps)=\ro_t
A_{i\mu}(-\eps)=0$, plugging $\ro_t E_i$ into \eqref{eq:ddA} and
integrating over a small interval $t\in [-\eps,+\eps]$ yields
\begin{align}
\ro_tA_{i\mu}(+\eps)& \approx \beta A_{i\mu}(0) \ln (1+s)
\end{align}
This and $E_i(\eps)=(1+s)E_i(0)$ are the initial conditions we start
with. In addition, we require $E,A,p\geq 0$ during the simulations.

\section{Application to European Sovereign Debt Crisis}

We apply our model to the GIIPS data mentioned above. Before looking at
the simulations of Eqs.~\eqref{eq:ddA}--\eqref{eq:ddE}, we estimate the
values of our parameters in the case of the GIIPS sovereign debt
crisis. 
\outNim{
{\color{red}
Note that our data is from the end of 2011. At that time two of
the most important holders of Greek debt, the National Bank of Greece
and Piraeus Bank, were already ``bankrupt'' in that they had negative
equity. We therefore exclude them in the simulations. We do this because
their trading behavior cannot be explained by our current model, not
because they lack importance.
}
}
\subsection{Estimating values of $\gamma=\alpha\beta$}

\begin{figure}
\centerline{
\includegraphics[width=8.7cm
]{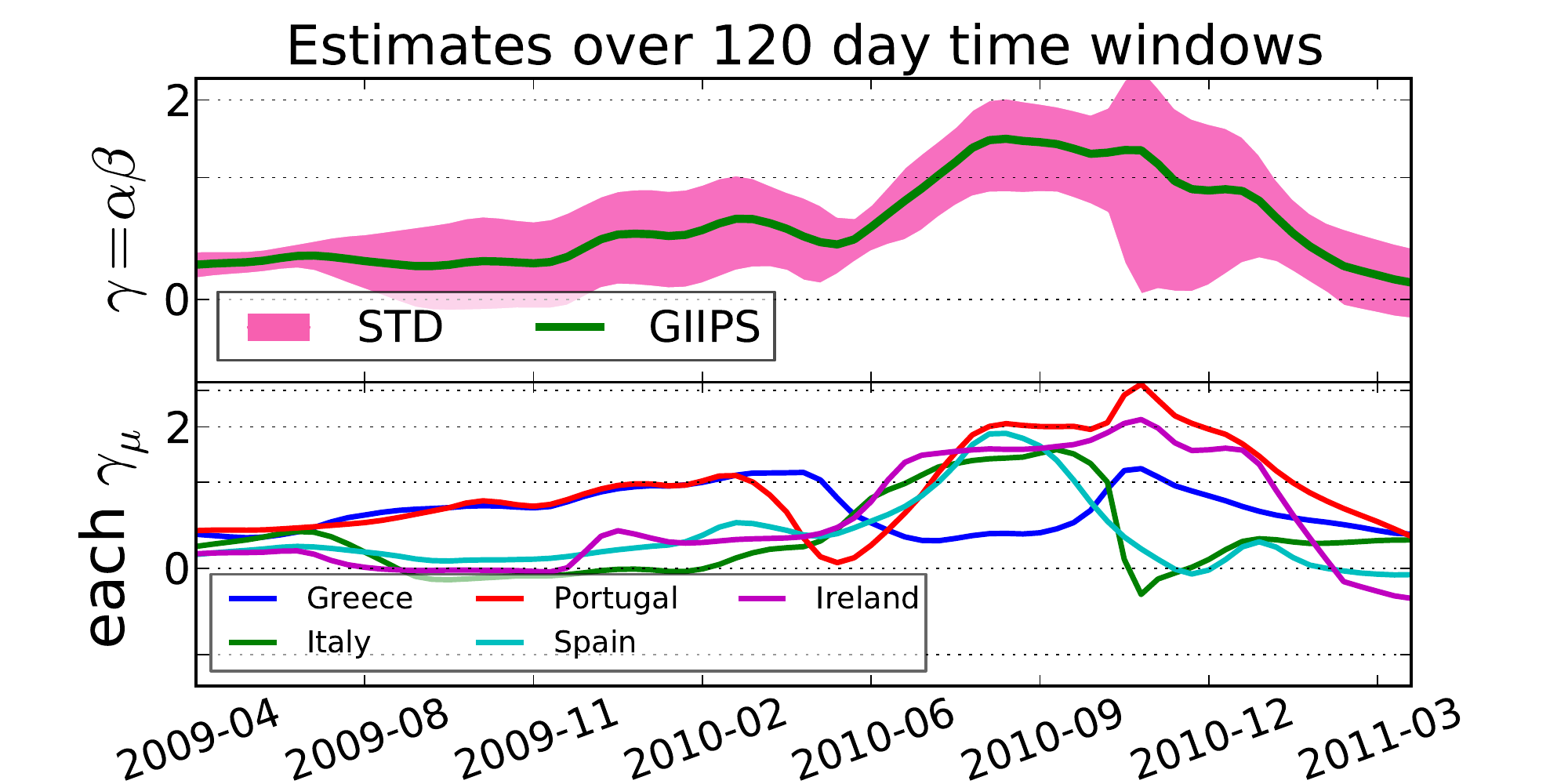}
\caption{ Estimates of $\gamma=\alpha\beta$ 
over 4 months
  periods. Top: the shaded purple region is the error-bars based
  standard deviation and the solid lines are the averages of different
  $\gamma$ calculated for each country. Bottom: Calculation of $\gamma_\mu$ for individual countries.  The fact that the values for
  different countries are close to each other is a sign that our
  assumption of ``herding'' (i.e. same $\alpha$ and $\beta$ for all
  GIIPS) is justified and that our model is applicable here.  As can be
  seen, before the height of the crisis $0<|\gamma|<1$ and then it
  gradually grows.  At the height of the crisis $\gamma\approx 2$.
  After the crisis we see $\gamma$ decrease again to $\gamma<1$.  Later
  we show that at $\gamma<1$ the system rolls into a new equilibrium,
  but when $\gamma>1$ the asset prices crash.  
Also note the timeline of bailouts: Greek bailout approved 2010/04 and 2010/09; Irish bailout 2010/10. these explain part of the movements in the lower plot.  
 The following stock
  tickers were used for each country (only the top 4 holders of each
  GIIPS for which stock prices were could be obtained from Yahoo!
  Finance): Greece: NBG, EUROB.AT, TPEIR.AT, ATE.AT; Italy: ISP.MI,
  UCG.MI, BMPS.MI, BNP.PA; Portugal: BCP.LS, BPI.LS, SAN; Spain: BBVA,
  SAN; Ireland: BIR.F, AIB.MU, BEN
\label{fig:gamma}}
}
\end{figure}

We use approximate versions of Eqs.~\eqref{eq:ddA}--\eqref{eq:ddE} to
estimate the product of parameters $\alpha$ and $\beta$ (details of the
approximation and the assumptions are discussed in the SI).  The
distribution of the assets is roughly log-normal, so a small number
banks hold a significant portion of each GIIPS country's debt. Thus
using only the equity of the dominant holders and denoting their sum by
$E^*_{(\mu)}$ for country $\mu$ will give us a good estimate of
$\gamma$. We estimate that the response times $\tau_A,\tau_B$ are at
most on the order of several days. Thus we will calculate
$\gamma=\alpha\beta$ over a period of four months to allow the system to
reach its new final state, and we can discard the second-order
derivatives.  This allows us to write
\[\delta A_{\mu} \approx \beta \sum_i {\delta E_i\over E_i
}A_{i\mu}\approx \beta {\delta E^*_{(\mu)}\over E^*_{(\mu)} }A^*_{\mu},
\] 
where $E^*_{(\mu)}$ denotes the equity of the dominant bank for asset
$\mu$. Using this approximation we can relate the first two equations\footnote{
The equity of the banks is mostly comprised of the shareholders' equity,
or common stocks. These banks usually have multiple stock tickers, but
there is generally one or two main stock tickers where most of the
equity is. We can use the movements in these main stocks to estimate
$\delta E^*_{(\mu)}/ E^*_{(\mu)}$. For this approximation we use the following formula:
\[{\delta E^*_{(\mu)}\over E^*_{(\mu)}} = {E_f-E_i\over (E_f+E_i)/2}\]
where $E_i$ is the stock price at the beginning of the period and $E_f$
is at the end of it.
},
\[{\delta p_\mu\over p_\mu} \approx \alpha {\delta A^*_\mu\over
  A^*_\mu}\approx \alpha \beta {\delta E^*_{(\mu)}\over E^*_{(\mu)}}.
\]
Thus we can approximate $\gamma$ as
\begin{equation}
\gamma \approx  \frac{\delta p_\mu/ p_\mu} { \delta E^*_{(\mu)}/ E^*_{(\mu)} }.
\end{equation}
We evaluate $\gamma$ for each country $\mu$. If the values are similar
for different $\mu$ values it may indicate that the ``herding effect''
is a factor. This both supports our model and suggests that it is
applicable to this problem. We evaluate $\gamma$ for the time period
between early 2009, when the crisis was just beginning, and early 2011,
when most government bailouts had either been paid or scheduled.

If we assume that the number of shares issued by our banks were roughly
constant over the period in question, the movements in stock prices may
be used as a proxy for the changes in the equity of banks. 
\outNim{120614

A plot of the
bond prices and stock prices of each country's top four debt holders
(which were obtained from Yahoo! Finance, list below figure) is given in
the SI in Fig.~\ref{fig:BondsE}, which shows that most of the time the
$\gamma$ values obtained from all the countries are similar.

}%
Many of the
major movements (or slope changes) in each country's $\gamma$ values
seem to coincide with bailout payment times.

Figure \ref{fig:gamma} shows the average $\gamma$ values during this
period with standard deviation error-bars. The bottom of the figure
gives estimates for $\alpha$ and $\beta$ assuming $|\alpha|=|\beta|$. A
more detailed plot with individual values of $\gamma$ obtained using
each country is also shown in Fig.~\ref{fig:gamma}.

Figure \ref{fig:gamma} shows that before the crisis $0<|\gamma|<1$ and
at its height $\gamma> 1$. Below we will explore the phase space in
terms of different values of $\alpha$ and $\beta$. An important finding
is that our model predicts that $\gamma>1$ is an unstable phase in which
a negative shock to the equity of any bank will cause most asset prices
to fall dramatically to nearly zero. Similarly, a positive shock will
cause the formation of bubbles.  When $0<\gamma<1$, on the other hand,
after a shock the system smoothly transitions into a new equilibrium
and, although some banks may fail, no asset prices will fall to zero.

\section{Simulations}

\begin{figure}
\centering
\includegraphics[width=3.3in]{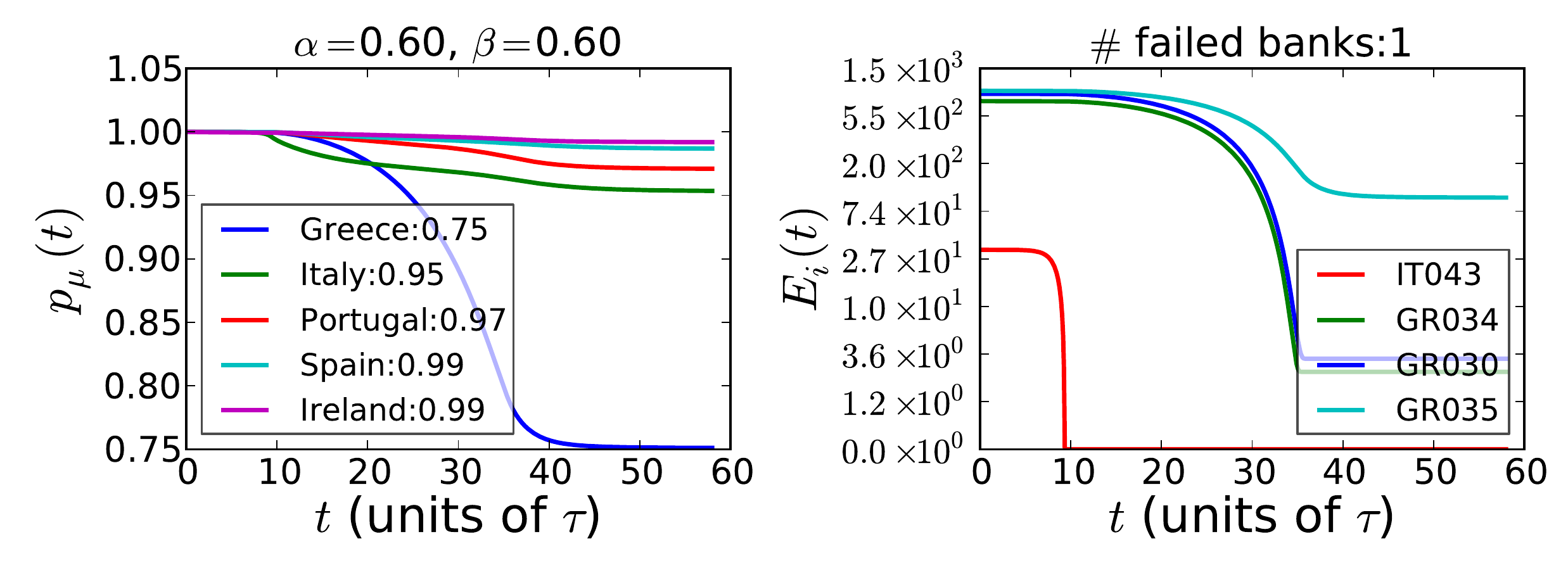}
\caption{Shocking ``Bank of America'' with $\alpha=\beta=0.6$. Left:
  plot of Asset prices over time. Greece incurs the greatest losses,
  falling to 75\% of original value. Final prices are listed in the
  legend. Right: Equities of the 4 ``most vulnerable banks'' (2 of major
  Greek holders incur large losses and one Italian bank is predicted to
  fail due to the shock). IT043 is Banco Popolare, which has very small
  equity but large Italian debt holdings. The next two are Agricultural
  Bank of Greece and EFG Eurobank Ergasias, which are among top 4 Greek
  holders.
\label{fig:shock-eq}}
\end{figure}

We find that when values of $\alpha$ and $\beta$ are small, e.g.,
$|\alpha \beta|<1$, shocking any of the banks in the network will
result in the same final state (see Fig.~\ref{fig:shk}). This is a new
stable equilibrium. If we shock the system a second time the prices do
not change significantly, i.e., less than 0.1\%).
Figure~\ref{fig:shock-eq} shows a sample of the time evolution of the
asset prices and the equity of the banks that incurred the largest
losses.

\begin{figure}
\centering
\includegraphics[width=3.3in]{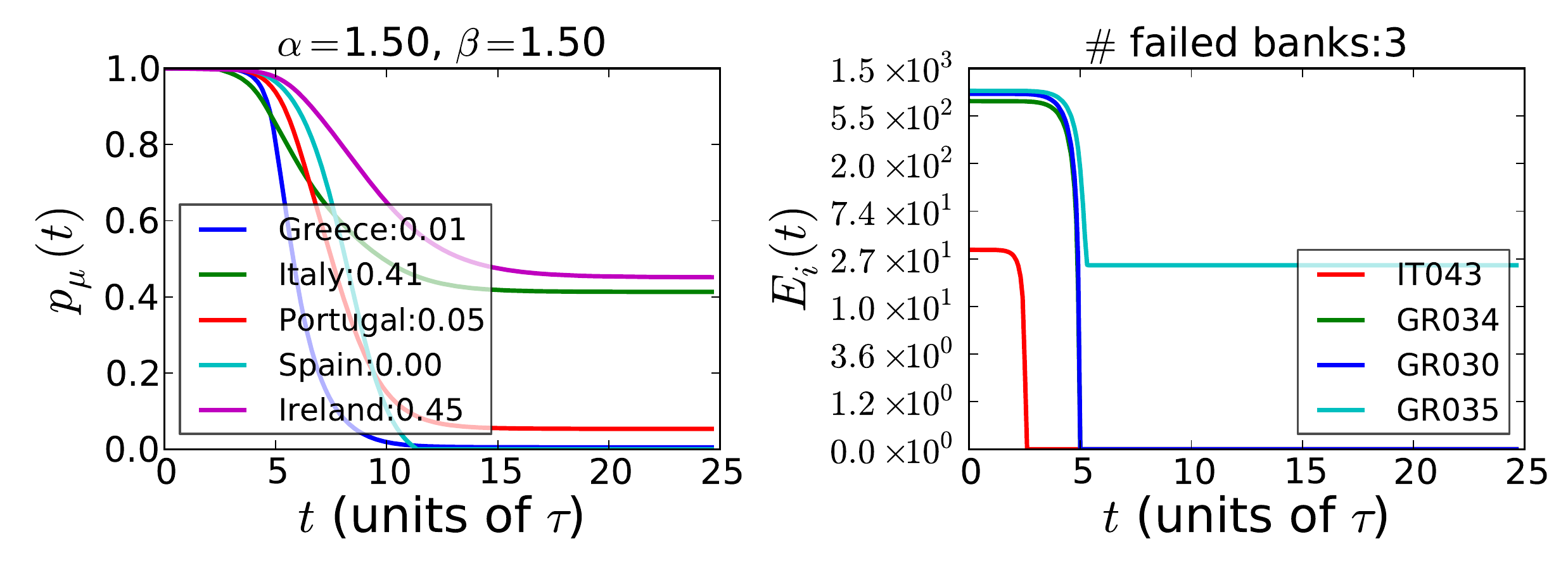}
\caption{Simulation for larger values of $\alpha$ and $\beta$ (values in
  legends are final price ratios $p_\mu(t_f)$). This time, in addition
  to Greek debt, Spanish and Portuguese debt show the next highest level
  of deterioration. The same four banks are the most vulnerable and this
  time two more of them fail. At $\alpha=\beta=1.5$ the damages are much
  more severe than at $\alpha=\beta=0.6$.
\label{fig:shock-hi}}
\end{figure}

Figure \ref{fig:shock-eq} shows results that seem in line with what
actually happened during the European debt crisis, although the damage
shown for Ireland is less than what actually occurred. In this figure,
bailouts are disregarded. Three of the four most vulnerable banks (MVB)
shown in Figs.~\ref{fig:shock-eq} and \ref{fig:shock-hi} are holders of
Greek debt. In this simulation, Greek debt is the asset that loses the
most value. Note that the loss prediction produced by the model is based
solely on the network of banks holding GIIPS sovereign debt and provides
information about the economies of these countries, with Greece
experiencing the largest loss, followed by Portugal (real-world data
indicates that Ireland's loss was as severe as Portugal's).

\begin{figure*}
\centering
\includegraphics[trim = 1.4in 0 1.2in .5in, clip,width=5.3in
]{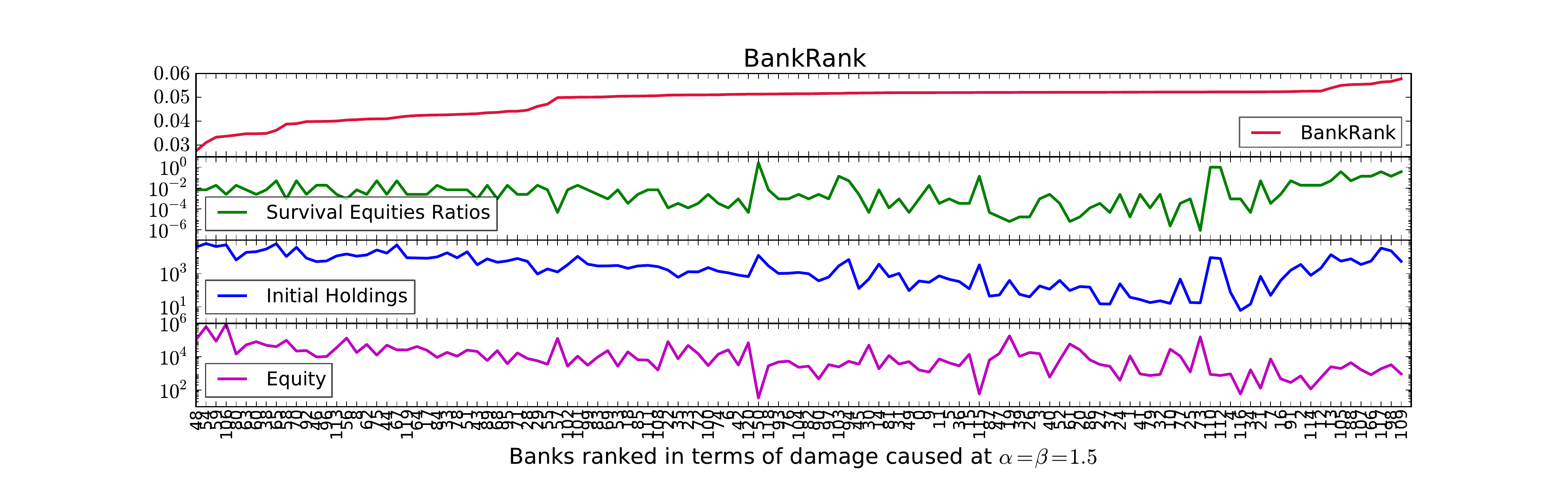}
\includegraphics[trim = .25in 0 .7in .7in,width=1.7in]{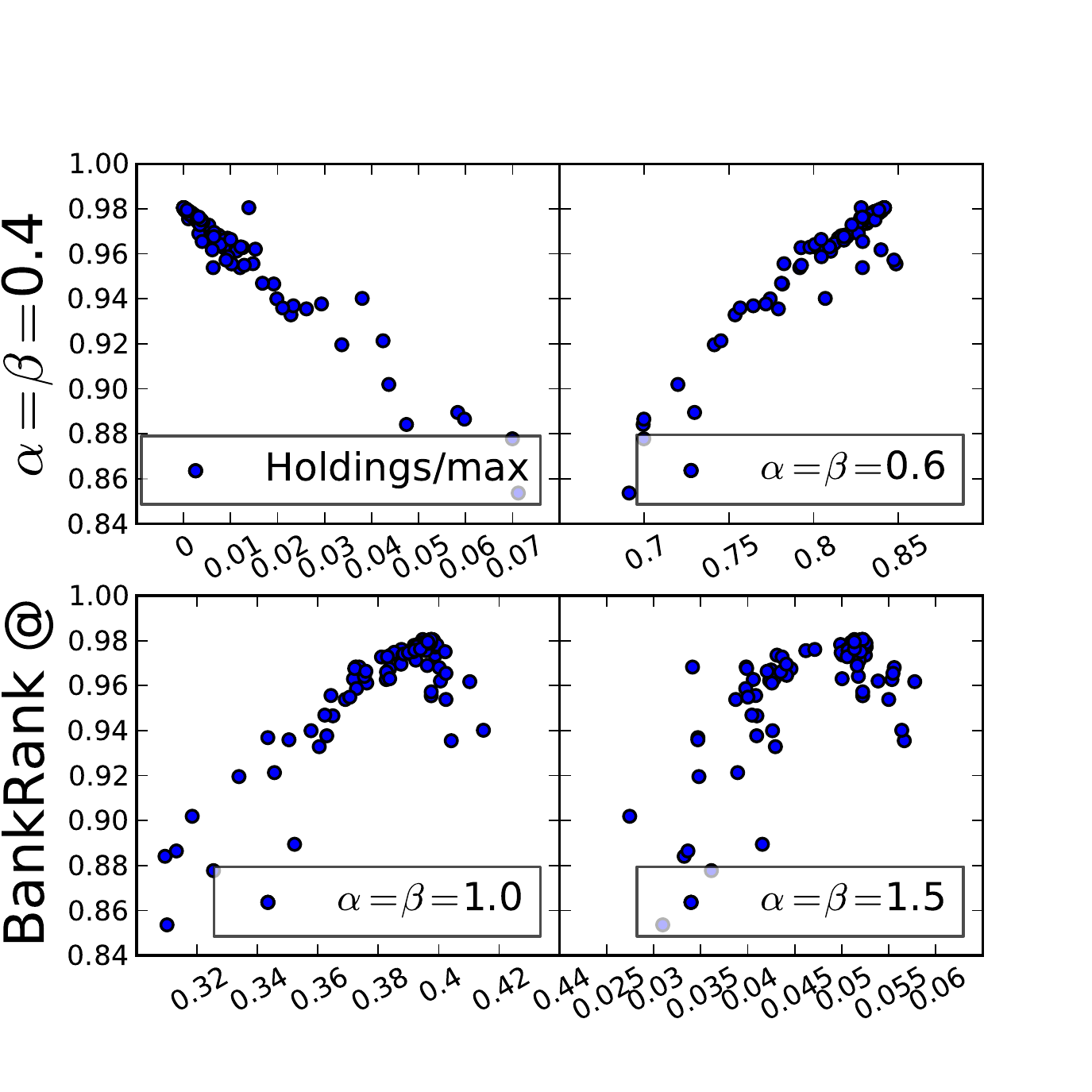}
\caption{ 
Left: Top: BankRank \eqref{eq:BankRank}: Ranking the banks in
  terms of the effect of their failure on the system. Top plot shows the
  ratio of final total GIIPS holdings in the system to the initial total
  GIIPS holdings. In this test we do two things. First we increase the
  equity of the 4 most vulnerable banks up to their total holdings
  ($E_i\to \sum_\mu A_{i\mu}$). Then, in each run, we adjust one bank's
  equity to just below the minimum equity that would allow it to survive
  the dynamics (the threshold is found empirically). The initial shock
  to the system is imposed through shocking the equity of a different
  bank (not the one we artificially brought to bankruptcy).  This way
  the BankRank tells us how much monetary damage the failing of one bank
  would cause.  The ranking changes slightly for different values of
  $\alpha $ and $\beta$. At low values,
  e.g. $\alpha=\beta<1$, the total assets completely determines the
  ranking. But at larger values, i.e. the ``unstable'' regime, the
  correlation reduces significantly. Right: a scatter plot of the BankRank at $\alpha=\beta=0.4$ (vertical axis) versus total holdings over maximum holding (Holding/max), and versus BankRank for other values of parameters at $\alpha=\beta=0.6,1.0, 1.5$. BankRank at $\gamma=\alpha\beta<1$ is directly anti-correlated with the initial holdings. BankRank at $\gamma=\alpha\beta>1$  isdifferes significantly from $\alpha=\beta=0.4$, which tells us that in the unstable regime $\gamma>1$ it is no longer true that only the largest holders have the highest systemic importance.
\label{fig:Rank3}}
\end{figure*}

Note that the new equilibrium depends on $\alpha$ and $\beta$. Returning
to the real data, Fig.~\ref{fig:gamma} shows that before the onset of
the crisis the system responds to a shock by achieving a new equilibrium
similar its initial equilibrium (behavior similar to that shown in
Fig.~\ref{fig:shock-eq}). At the height of the crisis, however, when
$\gamma=\alpha\beta \approx 2$, even a small shock can have a
devastating effect and precipitate a crisis (see
Fig.~\ref{fig:shock-hi}). Although many banks incur significant losses
when $\alpha$ and $\beta$ values are at their highest, the same four
banks fail.

In the SI we  show the effect of rewiring the banks who lend to each country, meaning we take $A_{i\mu}$ and take random permutations of index $i$ so that the equities of banks connected to each country changes randomly. Interestingly, such a rewiring changes the damages suffered by GIIPS bonds entirely, meaning that Greece will no longer be the most vulnerable.  
This shows that in our model, while qualitative behavior of the system only depends on $\alpha$ and $\beta$, the final prices and equities depend strongly on the network structure.

\section{Systemic Risk and BankRank}
\outNim{
{\color{red} NOT in SI!!
Figure \ref{fig:shk} (see the SI) shows a plot of the final prices that
result when individual banks are shocked. These final prices are very
similar.  Note that when $\alpha\beta>1$ most asset prices fall to zero
and thus cannot be used when comparing the systemic importance of
individual banks.
}
}
We thus take a different approach. We find that a bank can cause a large
amount of systemic damage when its equity level is at the bare minimum
necessary to survive a shock. Banks with very low equity fail rapidly,
no longer trade, and thus no longer transmit damage to the system. Banks
with equity sufficient to survive for a significant period of time, on
the other hand, continue to transmit damage into the system and thus
cause more damage than extremely weak banks. Based on this observation
we rank the banks using a ``survival equity ratio'' (SER), i.e., the
fraction of actual equity a bank needs in order to survive systemic
shock.  The total damage done to the system varies significantly from
bank to bank. To rank the systemic importance of each bank we measure
the effect their failure has on the system. Since normally no banks
other than the four mentioned above fail, we modify the data
slightly. The steps we take are as follows:

\begin{enumerate}

\item We increase the equity of the four failing banks to
  $E_i(0)=\sum_iA_{i\mu}(0)$ to keep them from failing and significantly
  damaging the system. Then when $\gamma=\alpha\beta<1$, the system
  becomes resilient to shocks and the drop in prices falls below 1\% (the
  system has reached a stable phase). When $\gamma>1$ the system
  continues to incur significant losses.

\item To assess the systemic importance of bank $i$, we run separate
  simulations with initial conditions changed to $E_i(0)\in [10^3
    E_i(0), 10^{-8}E_i(0)]$, until we find the threshold of survival
  under a small shock to any other bank $j$ ($i\ne j$).

\item We calculate the total GIIPS holdings $\sum_k (A\cdot p)_k$ left
  in the system. 
\end{enumerate}  
We define ``BankRank'' of $i$ to be the ratio of the final holdings to initial holdings if bank $i$ fails, i.e. BankRank of $i$ is equal to the amount of monetary damage the system would take if bank $i$ fails:
\outNim{
{
\color{red} WORDS!!! Given the initial conditions we are going to predict the future. In classical mechanics the synamics works like this,. Our approach is similar ABSTRACT!!! Laplacian determinism.
}
}
\begin{align}
\mbox{BankRank of~}i:R^i= {\sum_j (A\cdot p)_j(t_f)\over \sum_j (A\cdot
  p)_j(0)}\Bigg|_{\mathrm{Shocking\ } i}. 
\label{eq:BankRank}
\end{align}
The smaller the value of $R_i$, the greater the systemic importance of
bank $i$.

Fig. \ref{fig:Rank3} on the left shows the BankRank in the unstable regime at $\alpha=\beta=1.5$ and how it compares to the initial holdings, minimum ratio of equity required for survival, and initial equities. We observe some correlation between BankRank and each of these variables, but for many banks BankRank does not seem to follow any of these variables. On the right of Fig. \ref{fig:Rank3}, on the top left we see that BankRank in the stable regime at $\alpha=\beta=0.4$ has very high correlation with the initial holdings. In the other three plots on the right we see that BankRank changes significantly when we transition from the stable to the unstable regime. This again indicates that, while in the stable regime holding almost completely determine the systemic importance of a bank, in the unstable regime this is no longer the case and many small holders will have high systemic importance. 
\outNim{
Figure \ref{fig:Rank10} (see SI) shows how the failure of banks affects
individual bond prices $p_\mu$. The ten most destructive banks (and
their nationality) for each bond are listed, and in all cases, with the
exception of Ireland, the greatest damage is caused by banks that lend
to their own governments.
}

\section{Conclusion and Remarks}

We study the systemic importance of large institutional holders of GIIPS
sovereign debt and propose a simple, dynamic ``systemic risk
measurement,'' which we call BankRank.  We do not find any definitive
correlation between BankRank and diversification, but there is a strong
correlation between BankRank and total GIIPS holdingsq.  Our model
describes the response of the GIIPS system to shocks well enough to
reveal a ``herding effect,'' i.e., its presence is indicated whenever a
single value for the parameters can be used for the whole system. Our
methodology (i) can be used to model ``systemic risk propagation''
through a bi-partite network of banks and assets, i.e., it can serve as
a ``systemic stress testing'' tool for complex financial systems, and
(ii) it can be used to identify the ``state'' or ``phase'' in which a
financial network resides, i.e., if a banking system is in a fragile
state we can quantify it and determine how to transition the system into
a more stable state.

We suggest that our model could be useful as a monitoring and simulation
tool that allows policy makers to identify systemically important
financial institutions and to assess systemic risk build-up in the
financial network.

\begin{acknowledgments}

We thank the European Commission FET Open Project ``FOC'' 255987 and
``FOC-INCO'' 297149, NSF (Grant SES-1452061), ONR (Grant
N00014-09-1-0380, Grant N00014-12-1-0548),DTRA (Grant HDTRA-1-10-1-0014,
Grant HDTRA-1-09-1-0035), NSF (Grant CMMI 1125290), the European
MULTIPLEX and LINC projects for financial support.  We also thank
Stefano Battiston for useful discussions and providing us with part of
the data. The authors also wish to thank Matthias Randant and others for
helpful comments and discussions, and especially Fotios Siokis for
sharing important points about the data and the Eurozone crisis.

\end{acknowledgments}


\end{article}

\newpage
%
{\noindent\titlefont Supporting Information}
\setcounter{equation}{0}
\renewcommand{\theequation}{S\arabic{equation}}
\setcounter{figure}{0}
\makeatletter
\renewcommand{\thefigure}{S\@arabic\c@figure}
\makeatother

\section{Testing the Role of the Network}
Our goal is to determine how much of the above behavior is caused by the
network structure and how much by the value of the outstanding debts. To
examine the dependence of the results on network structure, i.e., to
determine which banks hold which country's debt and how much bank
equities matter, we randomize the network and redo our analysis. We do
not change the value of the total GIIPS sovereign debt held by the
banks. We only rewire the links in the network, changing the amount of
debt held by each bank and the countries to which each bank lends money.

\begin{figure*}[hb]
\centering
\includegraphics[width=3.3in]{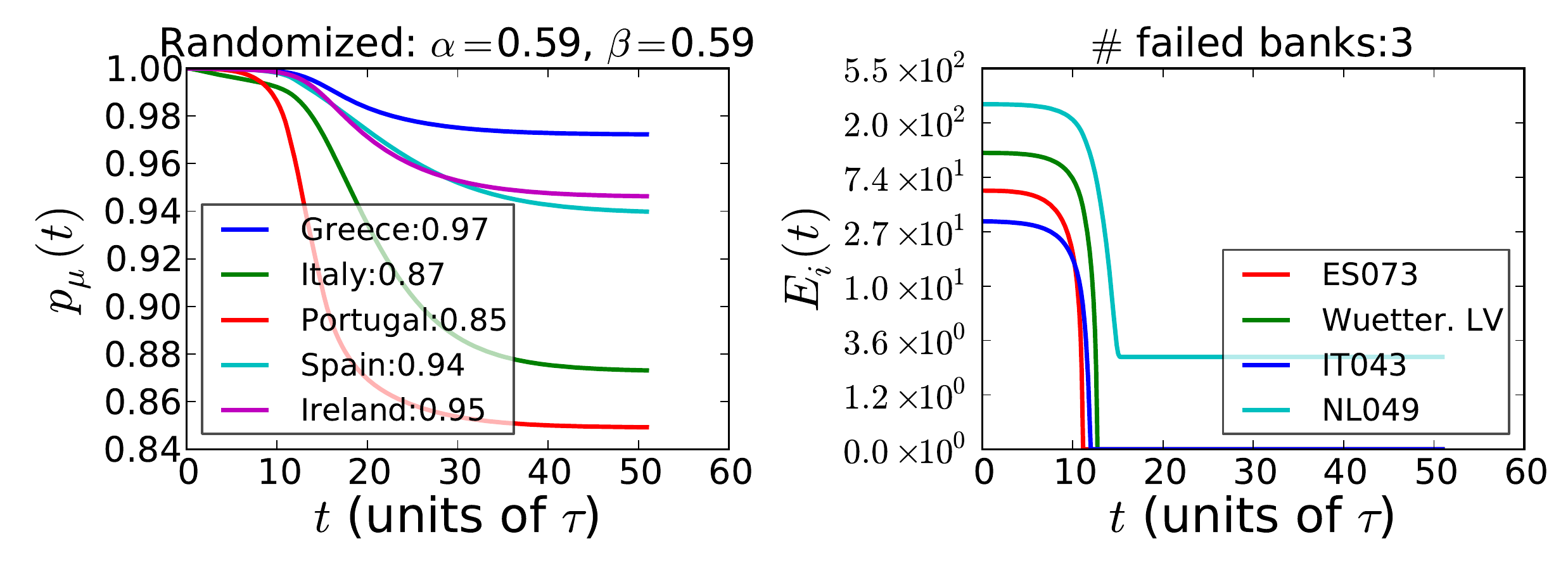}
\caption{Randomizing which bank lends to which country, while keeping
  total debt constant for each country. The results differ dramatically
  from the real world data used in Fig.~\ref{fig:shock-eq}. In this
  example Portugal and Italy lose the most value, while Greece is the
  least vulnerable. Other random realizations yield different results.
\label{fig:shk-rand}}
\end{figure*}
Figure \ref{fig:shk-rand} shows an example of this randomization and how
dramatically it changes the end result, and it demonstrates two
important features of the model: (i) system dynamics are strongly
affected by network structure, i.e., knowing such global variables as
the equity and exposure of individual banks is not sufficient, and (ii)
real-world data seems to indicate that it was the structure of the
network of lenders to Greece that caused Greek sovereign bonds to become
the most vulnerable.  This suggests that our model may be useful as a
stress testing tool for banking networks, or any network of investors
with shared portfolios.

\section{Shocking different banks}

\begin{figure*}[hb]
\centering
\includegraphics[width=3.3in]{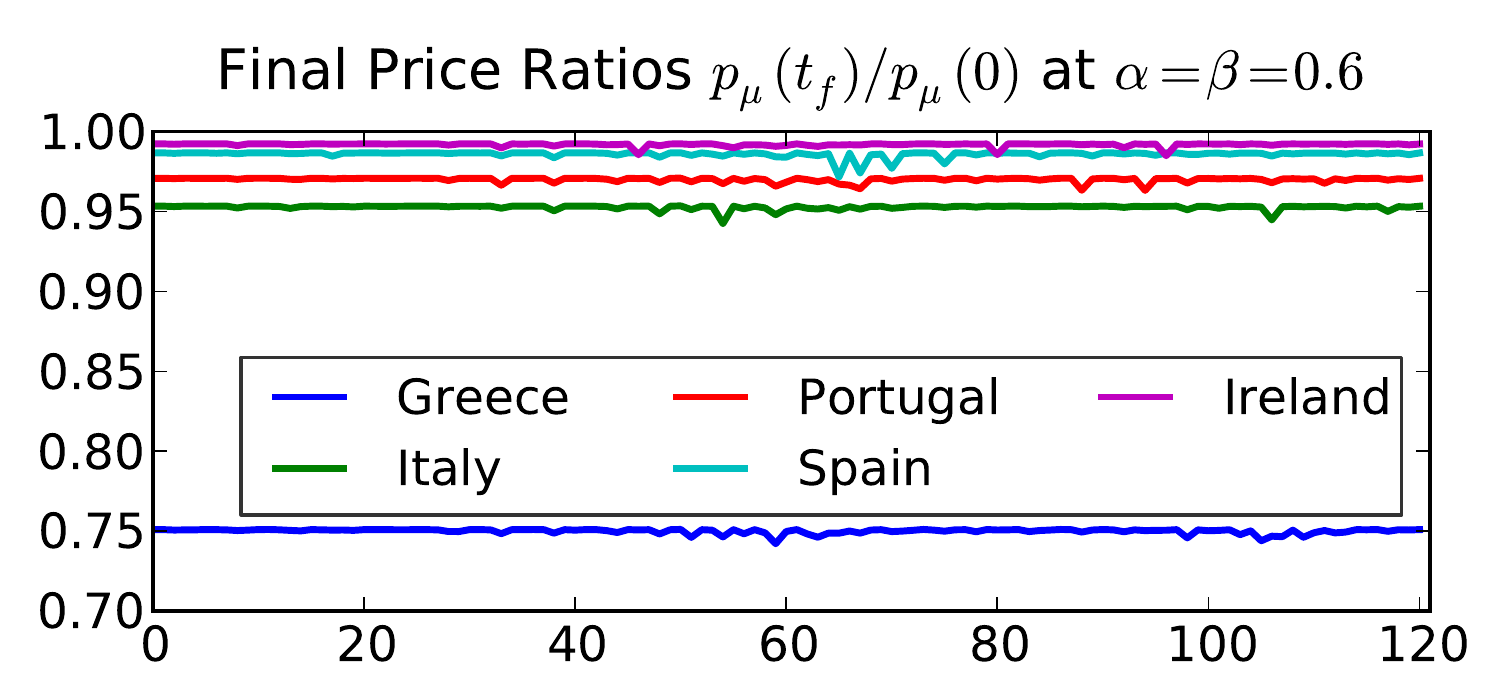}
\caption{Shocking different banks at $\alpha=\beta=0.6$. The final
  prices turn out very similar.
\label{fig:shk}}
\end{figure*}

Fig. \ref{fig:shk} shows the final prices found from shocking different
banks. They are all almost identical. However, the small variation and
the variations in the $A_{i\mu}(t_f)$ can be used to construct BankRank
and find that different banks have different mounts of influence.

\outNim{120714 

\section{Initial conditions for simulations}
We show later below that the choice of $\tau_{A,B}$ does not
affect the stability of the system and that the stability only depends
on $\alpha$, $\beta$ and the shock. The $f_i(t)$, which are changes in
the equity from what banks own outside of this network, can be thought
of as external noise or driving force. We use $f_i(t)$ to shock the
banks and make them go bankrupt. We shock a single bank, say bank $j$,
at a time by reducing its equity 10\% by putting $f_i(t)=-0.1E_j\delta_{ij}
\delta(t)$.\footnote{Note that the
  magnitude of the shock only rescales time, according to
  Eq.~\eqref{eq:ddE} because $f_i\to \lambda f_i$ is the same as
  $\ro_t\to \lambda^{-1}\ro_t$ and thus $\tau_{A,B} \to \lambda
  \tau_{A,B}$} \footnote{$\delta_{ij}$ is the kronecker delta, or the
  identity matrix elements, and $\delta(t)$ is the Dirac distribution or
  impulse function.} Starting with $\ro_t p_\mu(-\eps)=\ro_t
A_{i\mu}(-\eps)=0$, plugging $\ro_t E_i$ into \eqref{eq:ddA} and
integrating over a small interval $t\in [-\eps,+\eps]$ yields
\begin{align}
\ro_tA_{i\mu}(+\eps)& \approx \beta A_{i\mu}(0) \ln (1+f_i(0)/E_i(0))
\end{align}
This and $E_i(\eps)=0.9E_i(0)$ are the initial conditions we start
with. In addition, we require $E,A,p\geq 0$ during the simulations.

\section{Estimating $\gamma=\alpha\beta$}

From examining the behavior of the model for different values of
$\alpha$ and $\beta$ we found that the phases are roughly a function of
the product $\gamma=\alpha\beta$ and for $\alpha,\beta>0$, the curve
$\gamma=1$ seems to be approximately where the phase transition
happens. The $\gamma<1$ phase is the one where the system reaches a new
equilibrium without any of the prices collapsing to zero.

We now wish to know in which of the two phases the real GIIPS system
is. We will try to estimate the value of $\gamma=\alpha\beta$ using a
simplified version of the equations.

First we start by noting that the distribution of the holdings for each
country is roughly log-normal, or close to a power law, which means that
a handful of the institutions hold most of the debt. If we denote the
top holders holding by $A^*_\mu$ for each country we may approximate:
\[A_\mu \approx c\times A^*_\mu, \quad c\sim O(1)\]
Where the constant $c$ to correct for the contribution of other banks.
For each country we will only look at this dominant bank. The first
approximation is to assume:
\[\delta p_\mu(t+\tau_A)\approx \delta p_\mu(t),\quad \delta
A_{i\mu}(t+\tau_B)\approx \delta A_{i\mu}(t)\] 
We guess that the response time for both banks and the market are at
most of the order of a few days. For this approximation to be valid, we
examine changes over the course of several months.  

This allows us to
write:
\[\delta A_{\mu} \approx \beta \sum_i {\delta E_i\over E_i
}A_{i\mu}\approx \beta {\delta E^*_{(\mu)}\over E^*_{(\mu)} }A^*_{\mu}\] 
Where $E^*_{(\mu)}$ denotes the equity of the dominant bank for asset
$\mu$. With this approximation, we can relate the first two equations in
the following way:
\[{\delta p_\mu\over p_\mu} \approx \alpha {\delta A^*_\mu\over
  A^*_\mu}\approx \alpha \beta {\delta E^*_{(\mu)}\over E^*_{(\mu)} }\] 
Thus, we may be able to approximate $\gamma$ with:
\begin{equation}
\gamma \approx \frac{\delta p_\mu/ p_\mu} {\delta E^*_{(\mu)}/ E^*_{(\mu)} }.
\label{eq:gammaest}
\end{equation}

The equity of the banks is mostly comprised of the shareholders' equity,
or common stocks. These banks usually have multiple stock tickers, but
there is generally one or two main stock tickers where most of the
equity is. We can use the movements in these main stocks to estimate
$\delta E^*_{(\mu)}/ E^*_{(\mu)}$. 

For this approximation we use the following formula:
\[{\delta E^*_{(\mu)}\over E^*_{(\mu)}} = {E_f-E_i\over (E_f+E_i)/2}\]
where $E_i$ is the stock price at the beginning of the period and $E_f$
is at the end of it.

\outNim{
120614

Fig. \ref{fig:BondsE} shows a
snapshot of the main stocks for these banks from March 2009 to late 2012
from Yahoo! Finance. The prices are rescaled by dividing by the stock
price at the beginning of March 2009 (hence the coincidence of all
curves at the beginning).

\begin{figure*}[b]
\centering
\includegraphics[width=3.3in]{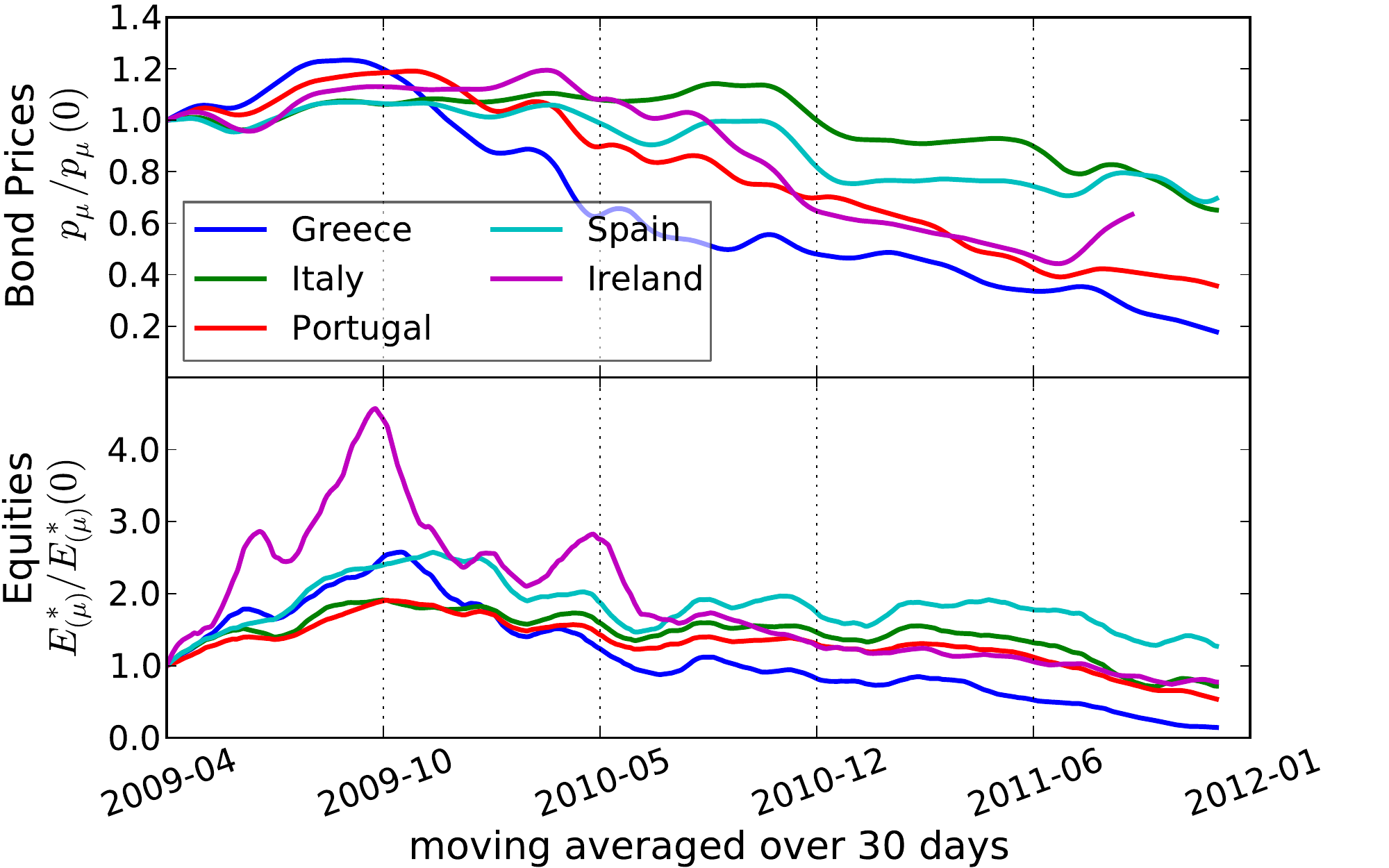}
\caption{Rescaled and 30 day moving average of bond prices (source
  Bloomberg) and stock prices (source Yahoo! Finance) of the dominant
  holders of GIIPS debt.
\label{fig:BondsE}}
\end{figure*}

Using stock prices from Yahoo! Finance, we are able to approximate
$\delta E^*_{(\mu)}/ E^*_{(\mu)}$ for parts of the time series where the
market trend is more or less constant. 


120614
}

\outNim{ 
\begin{table}[b]
\caption{ The estimated value of $\delta E^*_{(\mu)}/ E^*_{(\mu)}$ for
  three periods using the major tickers' stock prices of the dominant
  holders as a proxy for total equity.  } \centering {\tiny
\begin{tabular}{c|l|l|l|l|l}
Period & Greece & Italy & Portugal &Spain&Ireland \\
\hline
Mar,`09-Nov,`09& 1.69 & 0.90 &  0.90 &  1.46  &1.30\\
Nov,`09-Jul,`10 & -1.70 &-0.46 &-1.03 &-1. & -0.38 \\
 Jul,`10-Mar,`11& 1.29 &   0.96 & 3.56 & 1.94 & -1.72
\end{tabular}
}
\end{table}
}

\outNim{120614

An estimatesd value for $\gamma$ using each country is shown in
Fig. \ref{fig:GIIPSgamma}. As can be seen, in the first half of the
interval shown in the figure the values of $\gamma$ obtained from
different countries are consistent with each other. In the later
stages, which falls in the middle of the GIIPS debt crisis, the values
fluctuate a lot and the spread increases too.

\begin{figure*}
\centering 
\includegraphics[width=3.3in]{1114GIIPS_EBA_estimates_120.pdf}
\caption{An estimatesd value for $\gamma$ using each country.  While
  there are many fluctuations, the values of $\gamma$ obtained from
  different GIIPS countries stay close to each other. The individual
  movements may be attributed to certain events that affected a
  particular country. For instance, the dip in Greek and Portuguese
  $\gamma$ in mid 2010 is soon after first Greek bailout and the
  flattening out in late 2010 is around the Irish and Portuguese
  bailout. Eventually in 2011 $\gamma$ seems to go below 1 again, which
  according to our model means that the market calmed down and the
  dangerous unstable phase was over.
\label{fig:GIIPSgamma}}
\end{figure*}

120614
}

\section{Other Values of $\alpha$ and $\beta$ and the Phases}

The examples we plotted above were all from the $\alpha,\beta>0$
quadrant. This is what one normally expects from this system: $\beta>0$
means if a bank incurs a loss, they try to make up for it by making
money from selling GIIPS holdings; $\alpha>0$ means if there is selling
pressure (more supply than demand) the prices will go down. There are,
however, cases where the opposite happens. ``Contrarian'' agents in a
market are those who, for example, buy more GIIPS holdings when they
incur losses, hoping to recover some of their losses by reducing the
average cost of investment.  The market may also sometimes behave in a
contrarian fashion, when there is an anticipation of good news that
overcomes the selling pressure, or when other investors outside our
network (such as smaller investors or the European Central Bank) are
actually exerting a buying pressure. Plots of those cases can be found
in the SI in Fig. \ref{fig:contrarian}, where in general the numerical
solutions in the contrarian regime lead to the following conclusions:

\begin{enumerate}

\item In the third quadrant $\alpha<0,\beta<0$, where both investors and
  market are contrarian, losses are devastating. Many more banks fail
  for a small negative value for both $\alpha$ and $\beta$ and the asset
  prices quickly plummet down to zero.

\item The second and fourth quadrant where $\gamma=\alpha\beta <0$ are
  almost identical. No banks fail in these regimes, but also the amount
  of money lost or generated during the trading is negligible. This
  makes these regimes (either the investors or the market is contrarian,
  but not both) good for preventing failures, but they are very
  undesirable for profit making.

\end{enumerate}

\begin{figure*}
\centering
\includegraphics[width=3.3in,height=1.2in]{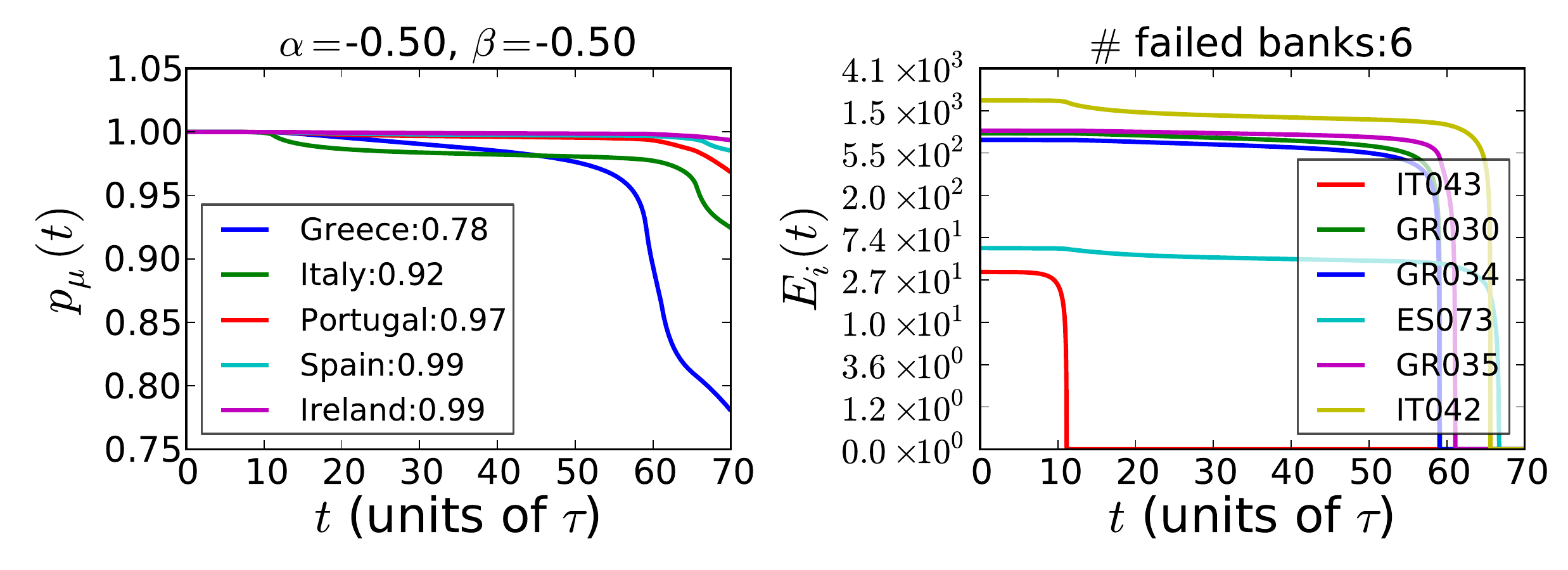}
\includegraphics[width=3.3in,height=1.2in]{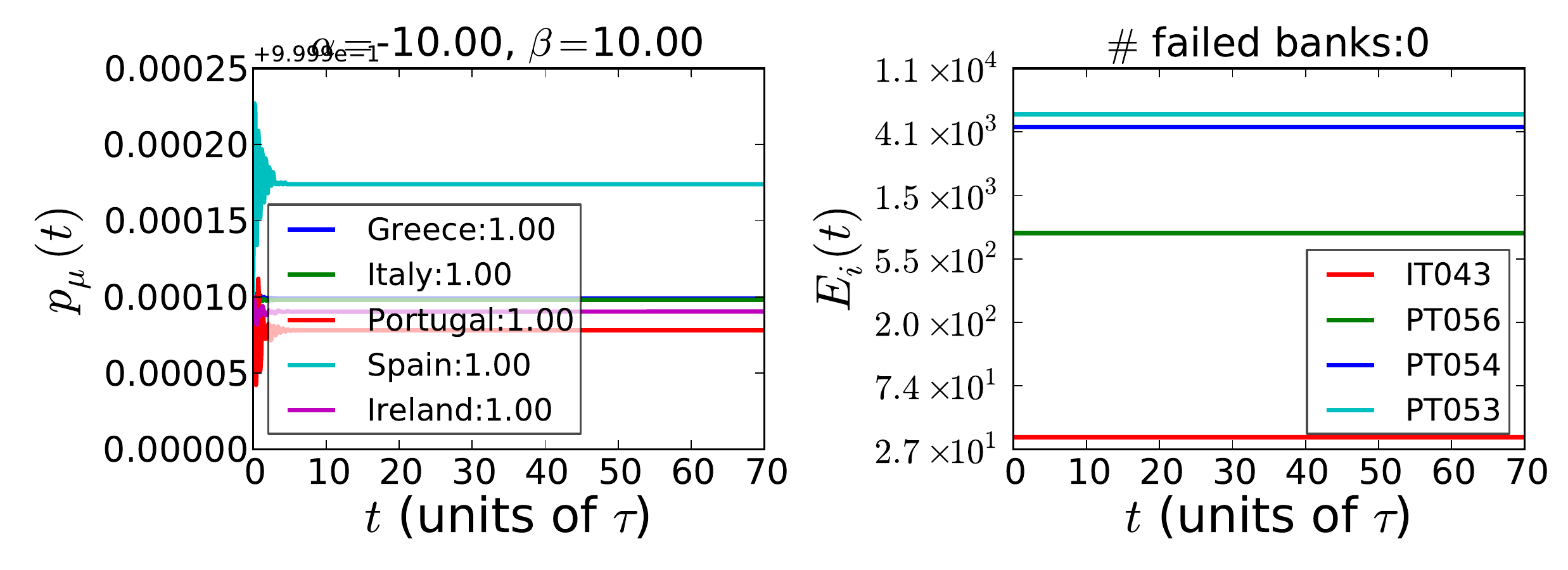}
\includegraphics[width=3.3in,height=1.2in]{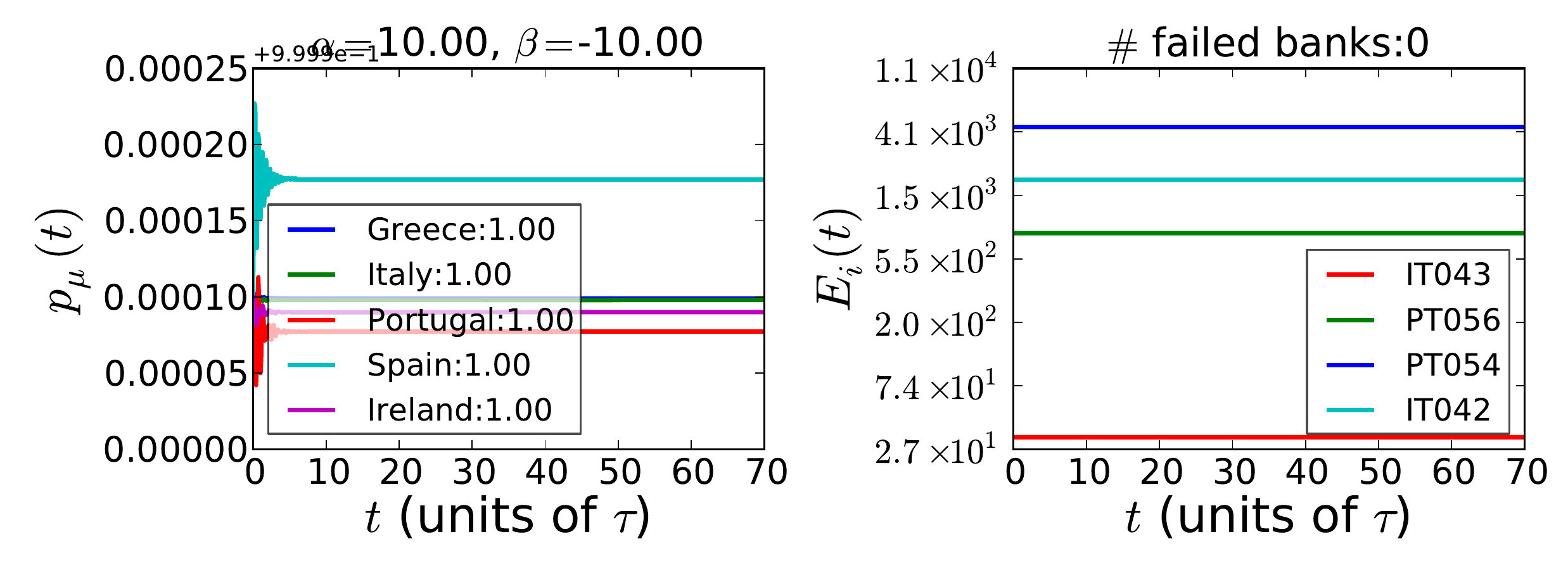}
\caption{Contrarian regimes: top, both $\alpha,\beta<0$. Here many banks
  fail, even for relatively small $\alpha, \beta$. The losses are
  devastating. Our model suggests that such a regime should be
  avoided. The bottom two plots show the two points $\alpha=\pm \beta$,
  $\beta=\pm 10$. The two results are almost identical. They also show
  that no appreciable amount of profit or loss is generated in these
  regimes, thus making them rather unfavorable for investors most of the
  time, but because of their safeness could be a contingency plan
  (buyout of bad assets by central banks is one such contrarian
  behavior).
\label{fig:contrarian}}
\end{figure*}

\outNim{120614

\section{Case of Ireland and Contrarian Behavior}

As mentioned above, looking into values of $\gamma=\alpha\beta$ during
the crisis reveals that the $\gamma$ obtained using major Irish debt
holders is very different from the rest of the countries (see SI
Fig. \ref{fig:GIIPSgamma}). Moreover, $\gamma$ from Ireland is negative,
while others are mostly positive. As it turns out, Franklin Templeton,
the dominant holder of Irish debt, is the cause of this difference. Our
source data shows that while many banks panicked and reduced their
exposure to the GIIPS debt, Franklin Templeton made a contrarian bid. It
aggressively {\em increased} its exposure to the Irish debt (an
additional 105\% by 12/2011), betting that Ireland would soon
recover. This bet was also partly secured by the already promised
bail-out, a manifestation of the ``moral hazard'' of the bail-outs.

120614
} 

\section{The phase Space}

\begin{figure*}
\centerline{
\includegraphics[trim = 1cm 0 20mm 0,  clip,
  width=1.7in]{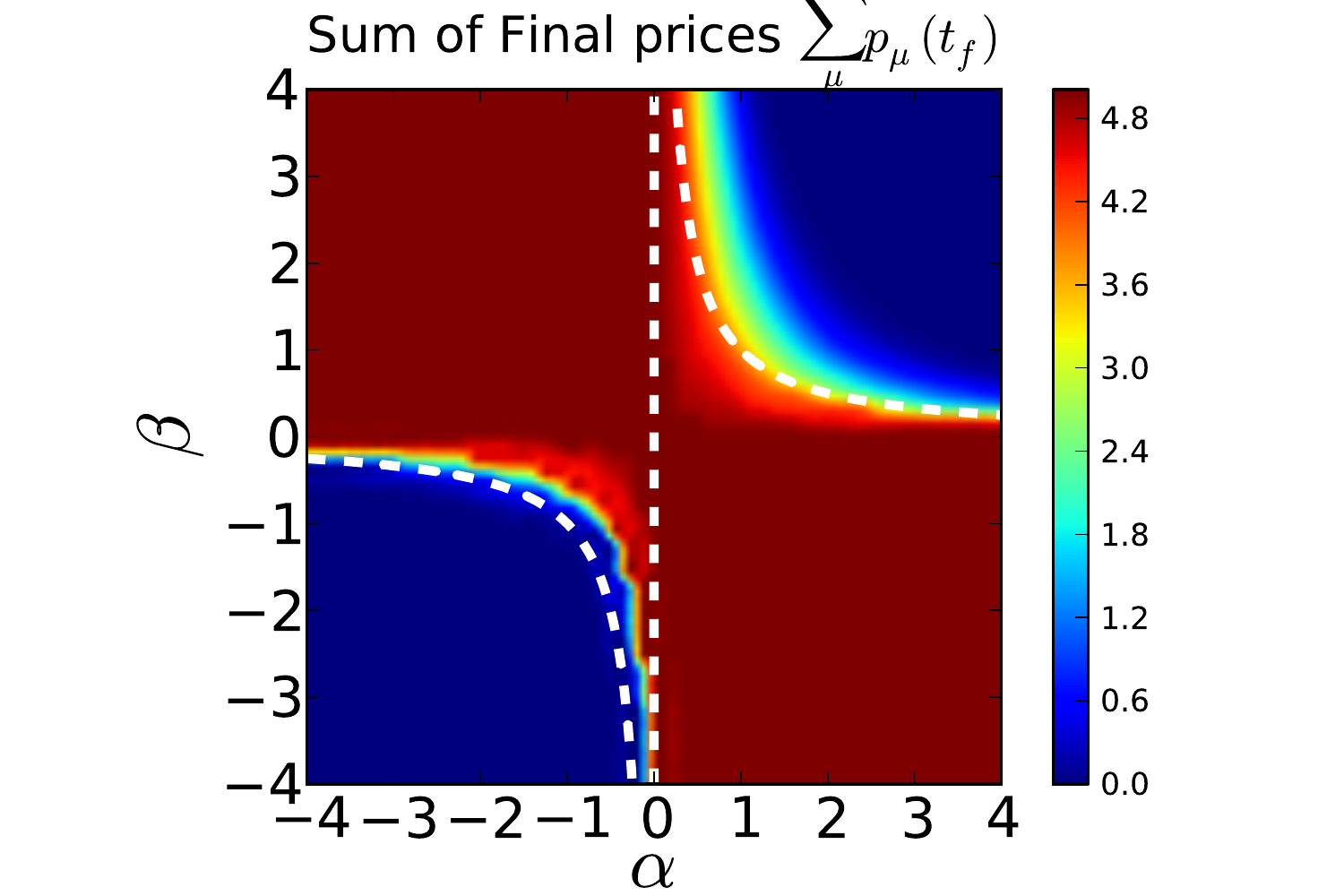}\includegraphics[trim = 1cm 0
  20mm 0, clip,width=1.7in]{phaseTimefit2601.pdf}} 
\caption{Left: Phase diagram of the GIIPS sovereign debt data, using the
  sum of the final price ratios as the order parameter.  We can see a
  clear change in the phase diagram from the red phase, where the
  average final price is high to the blue phase, where it drops to
  zero. The drop to the blue phase is more sudden in the $\alpha<0,
  \beta<0$ quadrant than the first quadrant. Right: The time it takes
  for the system to reach the new equilibrium phase. This relaxation
  time significantly increases around the transition region, which
  supports the idea that a phase transition (apparently second order)
  could be happening in the first and third quadrants.  The dashed white
  line shows the curve $\gamma=\alpha \beta =1$. It fits the red banks
  of long relaxation time very well. This may suggest that $\gamma=1$ is
  a critical value which separates two phases of the system.
 \label{fig:phase}}
\end{figure*}

Fig. \ref{fig:phase} shows an example of the average final prices and
relaxation time for the system for various values of $\alpha$ and
$\beta$. It seems the system has two prominent phases: One in which a
new equilibrium is reached without a significant depreciation in all of
the GIIPS holdings (upper left and lower right quadrants), and one where
all GIIPS holdings become worthless (above dashed line in the upper
right quadrant and all of lower left quadrant).  In the third quadrant
the transition is much more abrupt than in the first quadrant. In both
quadrants in the transition region the relaxation time becomes very
large, which means that the forces driving the dynamics become very
weak. Both the smoothness and the relaxation time growth seem to be
signalling the existence of a second order phase transition.  The phase
transition in the first quadrant seems to be described well by:
\[\gamma=\alpha \beta =1.\]
But this result is not exact and below we derive a more precise form for
this equation, which is:
\begin{equation}
\gamma = 1+ f_0,
\end{equation}
where $f_0$ is the magnitude of the initial shock $f_i(t)= f_0\delta(t)$
for a fixed $i$ that's being shocked. 
Analytical derivations of this are the subject of another paper which we are working on.

\outNim{
120614

We also simulated a 1 bank by 1
asset network (discussed in the SI) and observed a very similar shape
and position for the phase transition curve. 
In the SI we also
approximately derive the above relation for the transition curve
analytically, for certain regions of the first quadrant.

It is interesting to note that if one tries to write a Lagrangian for
such a system, the combination $\gamma=\alpha \beta$ is the only free
dimensionless parameter one may have in the most general Lagrangian
model, to lowest order in $E,A,p$ and at most two time derivatives. The
Euler-Lagrange equations of such a Lagrangian are not identical to our
equations, but are very similar to it. Specifically, the Lagrangian term
involving $\gamma$ is
\begin{equation}
L_\gamma=\gamma  \ro_t E^T A p - E^T A \ro_t p.
\label{eq:Lg}
\end{equation}

Thus $\gamma$ is the coefficient that adjusts the competition of two
effects: how much the loss of bank capital is affecting the market and
the loss of asset value affecting the banks.  When $\gamma=1$ the two
terms compete equally, and for other values of $\gamma$ one of the terms
dominates. The discussion of the Lagrangian model is beyond the scope of
this manuscript and will be the subject of another paper.

120614
}

\outNim{ 
\section{Remarks on the Euler-Lagrange Plus Dissipation Model}

Our model was proposed based on intuition about the system's response to
shocks.  A natural question to ask is: Is this system optimizing a type
of Lagrangian (``objective function'')? How can we formalize this
optimization?

The part of the Lagrangian in \eqref{eq:Lg} has no kinematics. Since the
kinetic terms are important, we investigate them in the SI, even though
they do not affect $\alpha$ and $\beta$.  Deriving the Euler-Lagrange
equations (see SI Eqs. \eqref{eq:dpmc}, \eqref{eq:dAmc} and
\eqref{eq:dEmc}) and comparing them to \eqref{eq:ddp} and \eqref{eq:ddA}
respectively, we find
\begin{equation}
\alpha ={-\gamma\over \gamma+1},\quad \beta= -(\gamma+1),\quad \beta = {-1 \over \alpha +1}
\label{eq:abg}
\end{equation}
which is consistent with the third equation in which we define
$\gamma\equiv \alpha \beta$, but this Lagrangian approach only allows
one free parameter, $\gamma$, instead of two.

This equation is significant because it shows the relationship in
bipartite markets between price sensitivity $\alpha$ and panic factor
$\beta$ and thus measures to what extent the system is isolated from
``the rest of the world.''  When this relation is not satisfied, we
conclude that ``the rest of the world'' has a significant effect on the
market. We argue in the SI that the effect of the rest of the world can
be viewed as ``dissipation'' and can be incorporated by adding friction
terms to the Euler-Lagrange equations. If we include these effects
through a single friction constant $\lambda$, the above equations are
modified to
\begin{align}
\alpha&={\gamma\over \lambda-\gamma-1},
& \beta&=\lambda-\gamma-1 \label{eq:lab}, & \beta&={\frac{-1}{\alpha +1}}(1-\lambda ).
\end{align}
Comparing Eqs.~\eqref{eq:lab} and \eqref{eq:abg} reveals a way of
quantifying the effect of the rest of the world.
} 

\outNim{
120614

\section{Non-network Measures and Determining the MVB }

We claimed that the most vulnerable banks (MVB) cannot be fully
identified without the dynamics, though for relatively calm situations,
the ranking of their equity compared to other banks is a fairly good
proxy. In other words, when $\alpha\beta$ is small (of order 1) all
failing banks have small equities compared to others, but not all banks
with small equities are vulnerable. This is intuitively clear because a
bank with small equity may not have invested in risky assets at all and
therefore be safe. This also shows that the network connections matter
in who will fail and who will not.  A second approach that comes to mind
immediately is that maybe the ratio of equity to GIIPS holdings is the
determining factor. However, This is also not the case and there is not
much correlation between the two.  It is true that these failing banks
again have small ratios $E_i/(Ap)_i$ but many banks with small values of
this ratio are ending up with big losses.

120614
}

\section{Initial conditions after the shock 
}

The $f_i(t)$, which are changes in the equity from what banks own
outside of this network, can be thought of as external noise or driving
force. We use $f_i(t)$ to shock the banks and make them go bankrupt. We
shock a single bank, say bank $j$, at a time by reducing its equity 10\%
by putting $f_i(t)=-0.1E_j\delta_{ij}
\delta(t)$ ($\delta_{ij}$ is the kronecker delta, or the
  identity matrix elements, and $\delta(t)$ is the Dirac distribution or
  impulse function). Note that the magnitude of the shock only rescales
  time, according to eq. \eqref{eq:ddE} because $f_i\to \lambda f_i$ is
  the same as $\ro_t\to \lambda^{-1}\ro_t$ and thus $\tau_{A,B} \to
  \lambda \tau_{A,B}$. In short, as we show below, starting with $\ro_t
p_\mu(-\eps)=\ro_t A_{i\mu}(-\eps)=0$, plugging $\ro_t E_i$ into
\eqref{eq:ddA} and integrating over a small interval $t\in
      [-\eps,+\eps]$ yields:
\begin{align}
\ro_tA_{i\mu}(+\eps)& \approx \beta A_{i\mu}(0) \ln (1+f_i(0)/E_i(0))
\end{align}
This and $E_i(\eps)=0.9E_i(0)$ are the initial conditions we start
with. In addition, we require $E,A,p\geq 0$ during the simulations.

With more details, for the numerical solutions we shock one bank, say
$i$, by imposing a 10\% loss on their equity
$$\delta E_j(t=0) = f_j(0) = \tilde{f}\delta_{ij} E_j(0)=-0.1\delta_{ij}
E_i(0)$$
We integrate equations \eqref{eq:ddA}-\eqref{eq:ddE} from shortly
before the shock $t=-\epsilon$ to shortly after it, $t=+\epsilon$ and
find that, if values of $E,A,p$ are finite, then $A,p$ will remain
continuous\footnote{In \eqref{eq:ddE}, $A$ cannot jump to infinity,
  because that would make \eqref{eq:ddp} wrong. Also, if $\ro_t p$
  absorbs the $\delta (t)$, integrating \eqref{eq:ddp} would require
  $A\propto \delta(t)$ again. Thus the only solution is to have $E$
  absorb the $f\delta(t)$, which means $\ro_t A,\ro_t p<\infty$ and
  thus $A,p$ both remain continuous.}. Then, integrating \eqref{eq:ddA}
with this assumption yields
\newpage
\begin{align}
\int_{-\epsilon}^\epsilon dt\left(\tau_B\ro^2_t A_{i\mu}+\ro_t  A_{i\mu}\right)&=\beta  \int_{-\epsilon}^\epsilon dt A_{i\mu}\ro_t\ln E_i \cr
\tau_B\ro_t A_{i\mu}(t)+A_{i\mu}(t)\Big|_{-\epsilon}^\epsilon&= \beta A_{i\mu}(0) \ln (1+\tilde{f})+O(\epsilon)\cr
\ro_tA_{i\mu}(+\epsilon)-\ro_tA_{i\mu}(-\epsilon)& \approx \beta A_{i\mu}(0) \ln (1+\tilde{f})
\end{align}
Which means that if we had started with $\ro_t A_{i\mu}(0) =0$ the
initial conditions can be changed to:
\begin{align}
E_i(0)&\to \tilde{f} E_i(0)
 & \ro_tA_{i\mu}(0)& \to \beta A_{i\mu}(0) {\ln (1+f_i)\over \tau_B}
\end{align}

\section{Robustness of the Ranking}

Fig. \ref{fig:RankMulti} compares the BankRank for three different
values of positive $\alpha$ and $\beta$. Some banks' BankRanks change
slightly, but the overall results are similar.

\begin{figure*}
\centering
{ \large BankRank} 
\includegraphics[width=7in, trim= 0 .2in 0 .6in, clip]{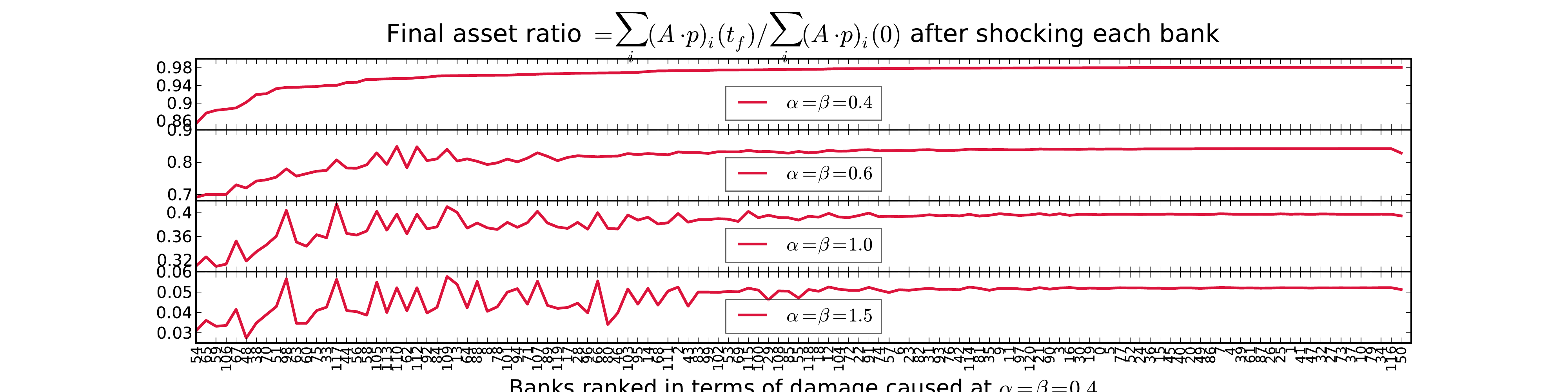}
\caption{BankRank for different values of $\alpha$ and $\beta$. After the phase transition to the unstable region (e.g. $\alpha=\beta=1.5$) the rankings change significantly.  
\label{fig:RankMulti}}
\end{figure*}

\begin{figure*}
\centering
\includegraphics[width=5.5in]{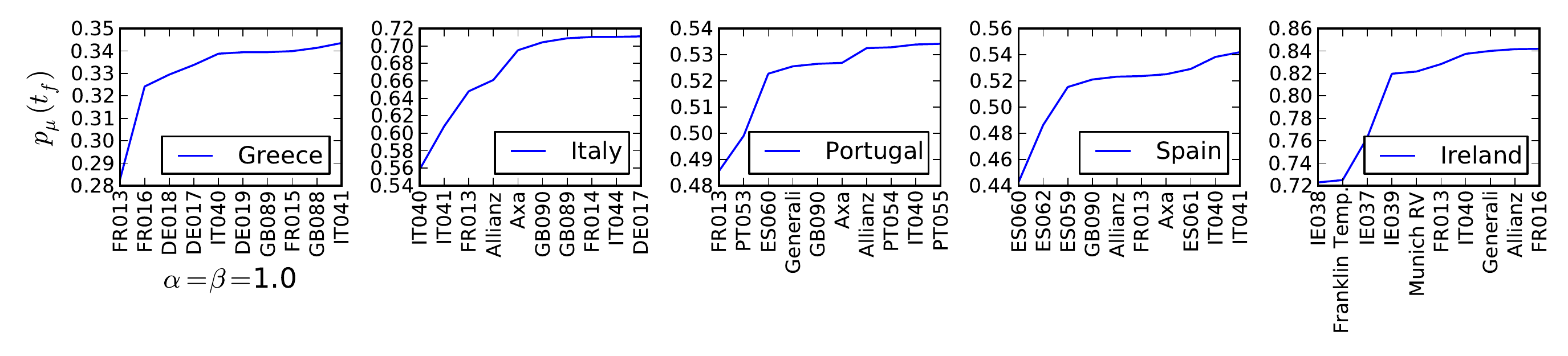}
\caption{Top 10 banks whose failure causes the most damage to the price
  of each country's sovereign bond. \label{fig:Rank10} }
\end{figure*}
Fig. \ref{fig:RankMulti} shows how the ranking changes as $\alpha$ and
$\beta $ increase. At small $\alpha\beta$ the ranking has haigh degree
of correlation with holdings, basically meaning that the larger the
money a bank holds, the more important it is. For large $\alpha\beta$,
however, this ranking changes significantly and some smaller players
become much more important than before. Fig.\ref{fig:Rank10} shows the
top 10 banks whose failure at $\alpha=\beta =1$ causes the largest damage
to each of the 5 GIIPS assets.

\section{BankRank and stability}

From examining the simulations more closely and from numerical analysis
of the differential equations \eqref{eq:ddp},\eqref{eq:ddA} and
\eqref{eq:ddE} in networks of few nodes, presented below, we see that as
expected the equations have either stable or unstable solutions. Stable
ones are those where the initial shock is dampened quickly and the
system goes to a new equilibrium, without any of the variables $E, A, p$
either collapsing exponentially to zero or blowing up
exponentially. Such behaviors in response to sudden rise or sudden fall
in $E$ in a 1 bank vs 1 asset system is shown in figure \ref{fig:1d}.

\begin{figure*}
\centering
\includegraphics[width=4.3cm]{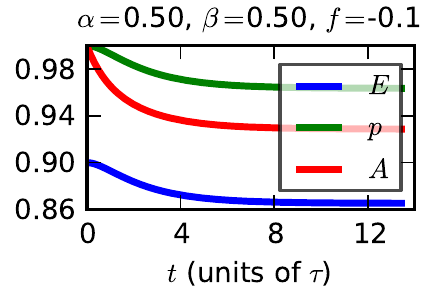}\includegraphics[width=4.3cm]{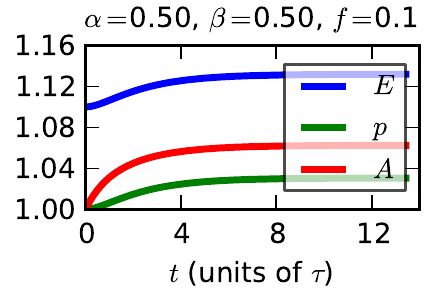}
\includegraphics[width=4.3cm]{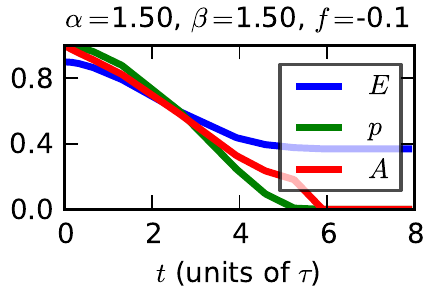}\includegraphics[width=4.3cm]{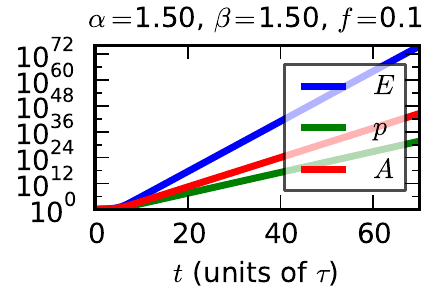}
\outNim{\includegraphics[width=1.7in]{1ddamped.pdf}\includegraphics[width=1.7in]{1d-damped.pdf}
\includegraphics[width=1.7in]{1ddiv.pdf}\includegraphics[width=1.7in]{1d-div.pdf}
}
\caption{Numerical solutions to the differential equations in a 1 bank
  vs 1 asset system. The upper plots show a ``stable'' regime, where
  after the shock none of the variables decays to zero or blows up, but
  rather asymptotes to a new set of values. The lower plots are in the
  ``unstable'' regime where positive or negative shocks either result in
  collapse or blowing up or collapsing of some variables.
\label{fig:1d}}
\end{figure*}

\outNim{ 
120614

A phase diagram
using $\ro_t E$ of the 1 by 1 system is shown in figure
\ref{fig:1dphase}.

\begin{figure*}
\centering
\includegraphics[width=1.5in]{Clar-1D.pdf}\includegraphics[width=1.5in]{Clar-1D-lin.pdf}
\caption{Phase diagram of the 1 bank vs 1 asset system, responding to
  sudden rise in $E$. Black denotes regions where $\ro_t E=A \ro_t p$
  was very large at late times, and light orange where it was close to
  zero. This is only plotting the $\alpha,\beta>0$ quadrant (left is a
  log-log plot, right is the regular linear scale diagram). The overlay
  are two fit functions for the phase transition curve. While
  $\alpha\beta=1$ is not a very good fit for large $\beta$ and small
  $\alpha$, it fits fairly well for large $\alpha$'s and we analytically
  prove this below.
\label{fig:1dphase}}
\end{figure*}

\section{Analytical results from the 1 Bank vs 1 Asset system}

Here we present the analytical solution to the 1 by 1 model and derive
the curve where the phase transition is happening in figure
\ref{fig:1dphase}. At any time $t$ the equations for a 1 by 1 system
become
\begin{align}
{(\ro_t+\tau_B\ro_t^2) A\over A}&=\beta {\ro_t E\over E}=\beta {A \ro_t p\over E}\cr
{(\ro_t+\tau_A\ro_t^2)p\over p}&=\alpha {\ro_t A\over A} \label{eq:1d1}
\end{align}
Below we will try to find the condition for a phase transition in the solutions to these equations.

120614
}

\outNim{

We can try to eliminate $A$ and $E$.  We first need to find the
expression for $\ro_t^2 A/A$ first. Taking another derivative from the
second equation yields
\begin{align}
{(\ro^2_t+\tau_A\ro_t^3)p\over p}-{(\ro_t+\tau_A\ro_t^2)p\ro_t p\over p^2}&=\alpha {\ro_t^2 A\over A}-\alpha \pa{{\ro_t A\over A}}^2 
\end{align}
combining this with the first equation results in:
\begin{align}
&{(\ro_t+\tau_A\ro_t^2)p\over p}+\tau_B
  {(\ro^2_t+\tau_A\ro_t^3)p\over p} =\gamma {A \ro_t p\over E }
  +O\pa{(\ro_t p)^2}\cr &\left[\tau_A\tau_B \ro_t^2+(\tau_A+\tau_B
    )\ro_t+\pa{1-\gamma {Ap\over E}} \right]\ro_t p = O\pa{(\ro_t
    p)^2}
\end{align}
Where the nonlinear term is again quadratic in $p$ (thus a genralized
form of the Fisher equation) and looks like
\begin{align}
O\pa{(\ro_t p)^2}=&\tau_B {(1+\tau_A\ro_t)\ro_t p\ro_t p\over p}\cr
&-\alpha \tau_B {\pa{(1+\tau_A\ro_t)\ro_t p}^2\over p} 
\end{align}
This time the dynamics is richer and we have a damped oscillator with a
driving force coupled to $p$ and nonlinearities of type $\sim (\ro_t
p)^2$. Taking the return $u \equiv \ro_t p$ as the fundamental
variable, the nonlinearities are roughly of type $u^2 + a \ro_t u^2$.
In short, the equations are
\begin{align}
\left[\tau \ro_t^2+\ro_t+\omega^2 \right]u  &=O\pa{u^2,\ro_t u^2}\cr
{1\over \tau}= {1\over \tau_B}+{1\over \tau_A}, 
\quad &\omega^2= {1-\gamma{Ap\over E}\over \tau_B+\tau_A},\cr{}
p(t)&= \int^t u(t') dt'
\end{align}
Although $\omega^2$ depends on $A,p$ and $E$, we can use an approximate
time dependent exponential ansatz $u\sim u_0 \exp[\lambda t]$. The
solutions to $\lambda$ are:
\[\lambda_\pm={-1\pm \sqrt{1- 4 \tau \omega^2}\over 2\tau}\]
When $\omega^2>0$ and $1-4 \tau \omega^2<0$ there will be oscillatory
solutions. For example when $\gamma{Ap\over E}<-1$, which only happens
for negative $\gamma$ we have such oscillatory solutions. This is
consistent with the simulations which showed the oscillatory behavior
was in the $\alpha\beta<0$ quadrants. For the stability, however we care
about the real solutions.

When $\omega^2<0$, which happens when $\gamma{Ap\over E}>1$, we will
have two real solutions with opposite signs. The presence of the
positive root signals an instability because the solution diverges. For
a delta function shock of magnitude $f$ at $t=0$ we found that:
\[E_0\to E_0(1+f)\]
Having initially scaled to $E_0=A_0=p_0=1$, the condition for existence of the positive root becomes:
\[t=0:\quad \gamma> {E\over Ap}= (1+f)\]
This dependence on the shock magnitude is normal, as a strong enough
kick can kick a prticle out of a local minimum.  The shock can be
arbitrarily small and therefore the absolute condition for stability is
as we anticipated
\[\mbox{unstable at:} \quad \gamma >1\]

Now the question is, which solution does the system pick when it is shocked. The return $\ro_t p$ is
\[\ro_t p(t) = u(t)= u_+ e^{\lambda_+ t}+u_- e^{\lambda_- t} \]
Since at $t=0$ the initial conditions dictated $\ro_t p(0)=0$ we have
\[u_+=-u_-\]
And therefore both solutions appear with equal strength. It follows that
whenever one of the solutions ($u_-$ in our case) is positive the
solution diverges. When $f>0$ a bubble forms and grows exponentially and
when $f<0$, because our variables are non-negative, the price just
crashes to zero.  This proves that the sufficient condition for
stability is $\gamma<1$. Also note that the nonlinear terms are all
proportional to $\ro_t p$ and therefore at $t=0$
\[O\pa{u^2(0),\ro_t u^2(0)}=0\]
and so the solution is exact at $t=0$.

\subsection{Validity of perturbation theory near the phase transition}

For the above solution to be valid we must confirm that the corrections
are small. We must find a small parameter that exists in the neglected
terms which allows perturbative solutions to be viable. We had two sets
of nonlinearities: 1) $O\pa{(\ro_t p)^2}$; 2) $\gamma Ap/E$. First let
us examine the nonlinear terms in $O\pa{(\ro_t p)^2}$.  Note that the
instability happens when the larger root $\lambda_-$ becomes
positive. Thus near the transition we have
\begin{align}
4\tau\omega^2&\ll1 \cr
\lambda_+&\approx -{1\over \tau} + \omega^2 \cr{}
\lambda_-&\approx - \omega^2
\end{align}
And so being close to the phase transition means $\lambda_-\ll
1/\tau$. The consequence of this is that for $O\pa{(\ro_t p)^2}$ we get
(using the $u_+=-u_-$ found above)
\begin{align}
\tau_A\ro_t)u=&\tau_A u_+\pa{\lambda_+ e^{\lambda_+ t}-\lambda_- e^{\lambda_- t}} \cr{}
\approx & \tau_A u_+\pa{\lambda_+ e^{\lambda_+ t}-\lambda_- e^{\lambda_- t}} \cr{}
O\pa{u^2}=&\tau_B {u(1+\tau_A\ro_t)u \over p}\cr &-\alpha \tau_B
{\pa{(1+\tau_A\ro_t)u}^2\over p}\cr 
\approx & \tau_B {u(1+\tau_A u_+(\lambda_+ -\lambda_-) )u \over p}\cr
&-\alpha \tau_B {\pa{(1+\tau_A\ro_t)u}^2\over p}\cr 
\end{align}

}
\outNim{ 120614

\section{Proof for $\gamma=1$ using properties of the phase transition}

Since we have coupled second order equations, the solutions may be
estimated using an exponential ansatz as follows. 
Equations \eqref{eq:ddA} and \eqref{eq:ddp} are decond order and therefore will naturally have two solutions for $A$ and $p$. Also, since $\ro_t E= A\ro_t p$, $E$ will also have two modes. Therefore the exponential ansatz must have at least two exponents. 
Thus for each of the three variables $X=E,p,A$ we have:
\[E\sim X_0 +X_1\exp[w_{X1} t]+X_2\exp[w_{X2} t]\]
%

In principle the exponents can be time-dependent, but we will first
try and see f there are asymptotically exponential solutions. Thus we
assume that they vary slowly with time. By choosing the units of $E,p,A$ to be such that at $t=0+\epsilon$, $p=A=1$ and the shocked equity is $E=1+f$, 
the boundary conditions that we had become:
\[
A_0=1-A_1-A_2,\quad p_0=1-p_1-p_2,\quad E_0=1+f-E_1-E_2\]
and:
\begin{align}
\ro_t E(0)&=0 =w_{E1}E_1+w_{E_2}E_2=0\cr
&= A\ro_t p= w_{p1}p_1+w_{p_2}p_2 \cr
\ro_t A(0)&= {\beta\over \tau_B} \ln (1+f)= w_{A1}A_1+w_{A_2}A_2
\end{align}
At the phase transition we expect the greater exponents, which we take to be $w_{X2}$, to become small relative to other time-sclaes in the problem, i.e. $\tau_A,\tau_B$, and change sign from negative (which would result in exponential decay) to positive (which results in divergence of $E,p,A$). This means that close to the phase transition:
\[|w_{X2}|\ll |w_{X1}|, \quad w_{X2}\ll {1\over \tau_A}+{1\over \tau_B}\]
From the initial conditions, this results in:
\begin{align}
|E_1|&= |{w_{E2}\over w_{E1}}E_2| \ll |E_2|,  &\Rightarrow E_0&=1+f-\pa{1-{w_{E2}\over w_{E1}} }E_2\approx 1+f-E_2\cr
|p_1|&\ll |p_2| ,&  \Rightarrow p_0 &\approx 1-p_2
\end{align}

For $A$ we have a little more details.
\[A_1= {\beta\over w_{A1}\tau_B} \ln (1+f)- {w_{A2}\over w_{A1}}A_2 \approx  {\beta\over w_{A1}\tau_B} \ln (1+f) 
\]
Which for small shocks $f\ll1$ reduces to:
\[A_1 \approx  {\beta\over w_{A1}\tau_B}f\]
Now back to the equations \eqref{eq:ddA}-\eqref{eq:ddE}. 
First let us reexamine the
third equation \eqref{eq:ddE}. The effect of a delta function shock
$f(t)=f_0\delta(t)$ is the above $\ro_t A$ and $E(+\epsilon)= (1+f_0)$. Since $|w_{X2}|\ll |w_{X1}$ and $w_{X1}<0$ we can neglect $\exp[w_{X1}t]$. The last equation becomes:
\begin{align}
\ro_tE&= w_{E2}E_2\pa{e^{w_{E2}t}-e^{w_{E1}t}}\approx w_{E2}E_2e^{w_{E2}t}\cr
&= A\ro_t p = \pa{1+A_1\pa{e^{w_{A1} t}-1}+A_2\pa{e^{w_{A2} t}-1}} w_{p2}p_2\pa{ e^{w_{p2} t}- e^{w_{p1} t}}\cr{}
&\approx \pa{1+{\beta\over w_{A1}\tau_B}f +A_2\pa{w_{A2} t}} w_{p2}p_2 e^{w_{p2} t}\cr{}
\end{align}
For arbitrary $t$ this relation can only hold if $w_{E2}=w_{p2}$. Thus we define:
\[w\equiv w_{E2}=w_{p2}\]

Let us also get an estimate for $w_{A1}$, the smaller exponent in $A$. We will go very close to the transition line where $w_{X2}\approx 0$. From Eq. \eqref{eq:ddA} we have:
\[{(\tau_B \ro_t +1)\ro_t A\over A}=\beta {\ro_t E\over E}\approx \beta {w E_2 e^{wt}\over E} \approx 0 \]
With an exponential ansatz the left hand side is:
\[(\tau_B w_A +1)w_A =0\]
The greater root is $w_{A2}=0$ and the smaller root is $w_{A1}=1/\tau_B$. Even away from the transition line we approximately have:
\[w_{A1}+w_{A2}\approx {1\over \tau_B}\]
Thus we can aproximate the expression for $A_1$ to:
\[A_1\approx {\beta\over \tau_B} \ln (1+f)\approx \beta f\]
This way Eq. \eqref{eq:ddE} becomes
\begin{align}
E_2&\approx \pa{1+\beta f +A_2\pa{w_{A2} t}} p_2 \approx (1+\beta f) p_2 
\end{align}
\outNim{ 
\begin{equation}
w_{E2}(1+f)=w_{p2} \label{eq:wewp}
\end{equation}
} 
Eq. \eqref{eq:ddA} becomes
\begin{align}
{(\tau_B w_{A2}+1)w_{A2}A_2 e^{w_{A2}t} \over 1+\beta f \pa{e^{t/\tau_B}-1}+A_2 w_{A2} t}&\approx  \beta {wE_2 e^{wt}\over 1+f + E_2 w t} 
\end{align}
Which again only holds if $w_{A2}=w$. Thus
\[w_{A2}=w_{E2}=w_{p2}=w\]
Again, note that the condition for being close to the transition point was:
\[w\ll {1\over \tau_A}+{1\over \tau_B}\]
Discarding higher than linear order terms in $w$ and looking at times $t/\tau_B\gg 1$ yields:
\begin{align}
{ A_2 \over 1-\beta f }&=  \beta { E_2 \over 1+f }+O(w) 
\end{align}
Performing the same procedure on Eq. \eqref{eq:ddp} results in (since $w\ll {1\over \tau_{A,B}}$)
\begin{align}
{(\tau_A w+1)wp_2 e^{wt}\over 1 + p_2 w t}&\approx  \alpha { wA_2 e^{w t} \over 1+\beta f \pa{e^{t/\tau_B}-1}+A_2 w_{A2} t}  \cr
p_2&=  \alpha { A_2  \over 1-\beta f }+O(w)= \alpha \beta {E_2\over (1+f)}+O(w)\cr{}
&\approx \gamma {1+\beta f\over 1+f} p_2 +O(w)
\end{align}
And so, the condition for the phase transition becomes:
\[\gamma = {1+f\over 1-\beta f}\]
\newpage
Now, taking the shock to zero $f\to 0$ results in a phase transition at:
\begin{align}
\mbox{Phase~Transition at:~}\gamma = 1
\end{align}

\outNim{

Plugging these into the equations \eqref{eq:ddA}-\eqref{eq:ddE} we get
($\tau_A=\tau_B=T$)
\begin{align}
Tw_A^2+w_A-\beta w_E &=0\cr
Tw_p^2+w_p-\alpha w_A &=0\cr
w_E\ln w_E- (w_p+w_A)\ln w_p&=0\cr
w_E(1+f)-w_p&=0
\label{eq:west}
\end{align}

Now, before going into the details of the solutions, we will give a
simple argument about why the phase transition should happen at
$\alpha\beta=1$. Because we have no discontinuity that can appear on our
equations, we expect the phase transition to be smooth and similar to a
second order transition (consistent with our observations from the GIIPS
system). This implies that near the phase transition the effective
potential is changing from having a stable minimum to having an unstable
maximum. It is like a second order potential is switching signs. The
result of this smooth transition from concave to convex is that at the
phase transition the potential becomes very flat, before switching to
being convex. For this reason in second-order phase transitions the
forces, which are the gradients of the potential, become very small and
as a result the relaxation time for the system to reach its final state
becomes very large (again, just as we observed in the phase diagram of
the GIIPS system).

We can use this fact about relaxation times being large to assume that
the acceleration terms in the first two equation of \eqref{eq:west} must
be small. In other words if we write $T_{A,p,E}\equiv 1/ w_{A,p,E}$
then \[T_{A,p,E}\ll T \] which is another way of saying the relaxation
time is large. Let's define the dimension-less rescaled exponents
\[\eps_{A,p,E}\equiv T w_{A,p,E}={T\over T_{A,p,E}},\]
and the condition for the phase transition curve becomes
\begin{equation}
\eps_{A,p,E}\ll1. \label{eq:wll1}
\end{equation}

Multiplying the first two lines of \eqref{eq:west} by $T$ results in
\begin{align}
\eps_A^2 + \eps_A -\beta \eps_E \approx  \eps_A -\beta \eps_E=0\cr
\eps_p^2 + \eps_p -\beta \eps_A \approx  \eps_p -\alpha \eps_A=0
\end{align}
Using the last equation of \eqref{eq:west} we have
$\eps_E=\eps_p/(1+f)$. Thus combining these two, the condition for phase
transition becomes:
\begin{align}
\eps_p =\alpha \eps_A &= {\alpha \beta \over 1+f} \eps_p\cr
\Rightarrow \gamma=\alpha\beta  &=(1+f)
\end{align}
This is the curve of the phase transitions for shock magnitude
$f$. Taking the external shock to zero $f\to0$ yields the position of
the phase transition
\[\mbox{Phase transition at:}\quad \gamma =1\]
Note that this argument also holds if $\tau_A\ne \tau_B$. At the phase
transition the relaxation time is larger than all time-cales involved so
$T_{A,P,E}\gg \tau_{A,B}$ and thus again the $w^2_{A,p,E}$ terms become
negligible and we get
\begin{align}
 w_A -\beta w_E=0\cr
w_p -\alpha w_A=0,
\end{align}
and again the same result is obtained by plugging in $w_E=w_p/(1+f)$.

}

120614
}

120714}

\section{The Banks}

The listed banks, insurance companies, and funds in the order of largest
holders of the GIIPS holdings is given in table \ref{tab:GIIPS}.

\newpage

\begin{table*}[t]
\centering
\caption{GIIPS debt data used in the analysis. All numbers are in
  million Euros. Our data is based on two sources: 1) The EBA 2011
  stress test data, which only includes exposure of European banks and
  funds (these are the ones where the ``Code Name'' is of the form
  CC123); 2) A list of top 50 global banks, insurance companies and
  funds with largest exposures to GIIPS debt by end of 2011 provided by
  S. Battiston et al. (These have a name as their ``Code Name''), which
  was consolidated by us.
\label{tab:GIIPS}}
\end{table*}

\newpage
{\footnotesize
\begin{longtable}{lllrrrrrrr}
\toprule
{} &                                        Name &           Code Name &  Holdings &   Equity &  Greece &   Italy &  Portugal &   Spain &  Ireland \\
\midrule
\endhead
\midrule
\multicolumn{3}{r}{{Continued on next page}} \\
\midrule
\endfoot

\bottomrule
\endlastfoot
0   &                                   GESPASTOR &       Gespastor &         3.5e+02 &  1.6e+03 &       0 &       0 &         0 & 3.5e+02 &        0 \\
1   &                                         M\&G &             M\&G &              37 &  1.1e+04 &       0 &       0 &        37 &       0 &        0 \\
2   &                            UNION INVESTMENT &      Union Inv. &         3.4e+03 &    7e+02 & 1.6e+02 &   2e+03 &        77 &   1e+03 &  1.5e+02 \\
3   &                                 ATTICA BANK &          Attica &         1.8e+02 &  1.5e+04 & 1.8e+02 &       0 &         0 &       0 &        0 \\
4   &                        MILANO ASSICURAZIONI &   Milano Assic. &              74 &  9.3e+02 &      23 &       0 &      0.71 &      49 &      2.1 \\
5   &                                    GROUPAMA &        Groupama &         4.4e+02 &  4.3e+03 &       0 & 4.2e+02 &        19 &       0 &        0 \\
6   &                                    AEGON NV &           Aegon &         1.1e+03 &  2.6e+04 &       2 &      65 &         9 & 9.8e+02 &       26 \\
7   &                                 RIVERSOURCE &    River Source &              48 &  7.4e+03 &      48 &       0 &         0 &       0 &        0 \\
8   &                                   AVIVA PLC &           Aviva &         1.1e+04 &  1.8e+04 & 1.5e+02 & 8.4e+03 &   2.3e+02 & 1.4e+03 &  7.2e+02 \\
9   &                               EMPORIKI BANK &        Emporiki &         2.9e+02 &  1.2e+03 & 2.9e+02 &       0 &         0 &       0 &        0 \\
10  &                               MELLON GLOBAL &          Mellon &              16 &  2.8e+04 &       0 &       0 &         0 &       0 &       16 \\
11  &                                       DAIWA &           Daiwa &         7.1e+02 &  7.3e+03 &       0 &   5e+02 &         0 &   2e+02 &        0 \\
12  &                                    FIDEURAM &        Fideuram &           2e+03 &  5.5e+02 &       0 &   2e+03 &         0 &       0 &        0 \\
13  &                                      UNIPOL &          Unipol &         1.3e+04 &  2.5e+03 &      26 & 1.2e+04 &   1.5e+02 & 1.1e+03 &  2.4e+02 \\
14  &                   WGZ BANK AG WESTDT. GENO. &           DE029 &         3.6e+03 &  1.9e+03 & 3.2e+02 & 1.4e+03 &   4.6e+02 & 1.2e+03 &  2.2e+02 \\
15  &                                  JYSKE BANK &           DK009 &         1.2e+02 &  1.4e+04 &      64 &       0 &        19 &      15 &       22 \\
16  &               OESTERREICHISCHE VOLKSBANK AG &           AT003 &         3.7e+02 &  4.8e+02 & 1.1e+02 & 1.5e+02 &        29 &      66 &       13 \\
17  &                              CAIXA PORTUGAL &      Caixa (PT) &         8.1e+03 &  2.4e+04 &      35 & 4.6e+02 &        30 & 7.5e+03 &       44 \\
18  &                                   BLACKROCK &       Blackrock &           2e+03 &    2e+04 & 1.2e+02 & 1.1e+03 &        29 & 7.1e+02 &       30 \\
19  &                             BANK OF AMERICA &            BofA &         3.8e+02 &  1.8e+05 &      13 & 2.5e+02 &       5.4 &      83 &       29 \\
20  &                       NORDEA BANK AB (PUBL) &           SE084 &         1.6e+02 &  2.6e+04 &       0 &      97 &         0 &      64 &      1.4 \\
21  &                      CAJA DE AHORROS Y M.P. &           ES077 &         1.5e+03 &     -&       0 &       0 &         0 & 1.5e+03 &        0 \\
22  &                              SELLA GESTIONI &           Sella &         6.6e+02 &  1.3e+02 &       0 & 6.6e+02 &         0 &       0 &        0 \\
23  &                              MITSUBISHI UFJ &      Mitsubishi &         1.6e+03 &  8.1e+04 &       0 & 9.2e+02 &        71 & 5.2e+02 &       62 \\
24  &                                         UBS &             UBS &         1.3e+03 &  4.8e+04 &      53 & 6.8e+02 &        55 & 4.4e+02 &       42 \\
25  &                                 OPPENHEIMER &     Oppenheimer &         2.4e+02 &  3.8e+02 &      15 &       0 &         0 & 2.2e+02 &        0 \\
26  &                                    VONTOBEL &        Vontobel &              18 &  1.2e+03 &      18 &       0 &         0 &       0 &        0 \\
27  &                                      NOMURA &          Nomura &              39 &  1.8e+04 &       0 &       0 &        20 &       0 &       19 \\
28  &                                   MACKENZIE &       MacKenzie &              15 &  3.4e+03 &      15 &       0 &         0 &       0 &        0 \\
29  &                                       AGEAS &           Ageas &         5.3e+03 &  7.8e+03 & 6.4e+02 &   2e+03 &     1e+03 & 1.1e+03 &  5.1e+02 \\
30  &                           DEUTSCHE POSTBANK &     De.Postbank &         9.2e+02 &  5.7e+03 & 9.2e+02 &       0 &         0 &       0 &        0 \\
31  &                              MORGAN STANLEY &     Morgan Sta. &         4.6e+02 &    5e+04 &       0 & 4.6e+02 &         0 &       0 &        0 \\
32  &                            HELVETIA HOLDING &        Helvetia &           1e+03 &  3.6e+03 &     7.6 & 7.2e+02 &        18 & 2.4e+02 &       15 \\
33  &                                   HWANG-DBS &           Hwang &              23 &  8.7e+02 &      23 &       0 &         0 &       0 &        0 \\
34  &                      ASSICURAZIONI GENERALI &        Generali &         1.7e+04 &  1.8e+04 & 1.3e+03 & 5.4e+03 &   3.1e+03 & 5.7e+03 &  1.7e+03 \\
35  &                                   AMLIN PLC &           Amlin &              15 &  1.6e+03 &       0 &       0 &         0 &      15 &        0 \\
36  &                          SWISS LIFE HOLDING &      Swiss Life &         5.9e+02 &  7.5e+03 &      30 & 1.7e+02 &        77 & 1.8e+02 &  1.3e+02 \\
37  &                               PHOENIX GROUP &         Phoenix &         3.2e+02 &  2.8e+03 &       0 & 2.3e+02 &        11 &      76 &      2.2 \\
38  &                                PRICE T ROWE &         PT Rowe &              15 &  2.6e+03 &       0 &       0 &         0 &       0 &       15 \\
39  &                                         AXA &             Axa &         2.9e+04 &  4.9e+04 & 7.6e+02 & 1.7e+04 &   1.5e+03 & 9.4e+03 &  7.5e+02 \\
40  &                                TOKIO MARINE &    Tokio Marine &              56 &    1e+04 &       0 &       0 &        30 &       0 &       26 \\
41  &                                  ROTHSCHILD &      Rothschild &         1.1e+02 &    6e+02 &      61 &       0 &        52 &       0 &        0 \\
42  &                               TT ELTA AEDAK &   TT Elta Aedak &              27 &  9.3e+02 &      27 &       0 &         0 &       0 &        0 \\
43  &                                     BALOISE &         Baloise &         7.9e+02 &  3.2e+03 &      84 & 2.7e+02 &        98 & 2.3e+02 &  1.1e+02 \\
44  &                                     NATIXIS &         Netaxis &         3.3e+03 &  2.1e+04 & 4.3e+02 & 1.3e+03 &   3.9e+02 & 8.6e+02 &  3.9e+02 \\
45  &                             CREDIT AGRICOLE &           FR014 &         1.7e+04 &  4.9e+04 & 6.6e+02 & 1.1e+04 &   1.2e+03 & 3.9e+03 &  1.6e+02 \\
46  &                                 JULIUS BAER &       Jul. Baer &         1.2e+02 &  3.5e+03 &      68 &       0 &         0 &       0 &       57 \\
47  &                          FRANKLIN TEMPLETON &  Franklin Temp. &         5.1e+03 &  9.7e+03 &       0 &       0 &         0 &       0 &  5.1e+03 \\
48  &            NOVA LJUBLJANSKA BANKA &           SI057 &         1.7e+02 &     -&      20 &      96 &        15 &      26 &       15 \\
49  &                                STATE STREET &       State St. &              51 &  1.6e+04 &       0 &       0 &        27 &       0 &       24 \\
50  &                                     ALLIANZ &         Allianz &         3.8e+04 &    1e+05 & 6.2e+02 & 2.9e+04 &   7.5e+02 & 7.1e+03 &  4.9e+02 \\
51  &                            VIENNA INSURANCE &          Vienna &              93 &    5e+03 &      21 &      13 &         0 &       7 &       52 \\
52  &                       BANCO POPOLARE - S.C. &           IT043 &         1.2e+04 &       33 &      87 & 1.2e+04 &         0 &   2e+02 &        0 \\
53  &                              COMMERZBANK AG &           DE018 &           2e+04 &  2.5e+04 & 3.1e+03 & 1.2e+04 &   9.9e+02 &   4e+03 &       32 \\
54  &                             LEGAL \& GENERAL &             L\&G &         3.8e+02 &  6.3e+03 &     1.1 & 3.3e+02 &       6.6 &      35 &      4.4 \\
55  &                                    EFFIBANK &           ES063 &           3e+03 &  2.7e+03 &      37 &       0 &        16 & 2.9e+03 &        0 \\
56  &                       INTESA SANPAOLO S.P.A &           IT040 &         6.2e+04 &  6.4e+05 & 6.2e+02 &   6e+04 &        73 & 8.1e+02 &  1.1e+02 \\
57  &                    IRISH LIFE AND PERMANENT &           IE039 &         1.9e+03 &  3.5e+03 &       0 &       0 &         0 &       0 &  1.9e+03 \\
58  &                           HSBC HOLDINGS PLC &           GB089 &         1.5e+04 &  1.3e+05 & 1.3e+03 & 9.9e+03 &     1e+03 &   2e+03 &  2.9e+02 \\
59  &                                 DANSKE BANK &           DK008 &         1.2e+03 &  1.3e+05 &       1 & 5.8e+02 &   1.1e+02 & 1.2e+02 &  4.1e+02 \\
60  &                ROYAL BANK OF SCOTLAND &           GB088 &           1e+04 &  9.6e+04 & 1.2e+03 &   7e+03 &   2.9e+02 & 1.5e+03 &  4.5e+02 \\
61  &                                 BNP PARIBAS &           FR013 &         4.1e+04 &  8.6e+04 & 5.2e+03 & 2.8e+04 &   2.3e+03 &   5e+03 &  6.3e+02 \\
62  &                                BARCLAYS PLC &           GB090 &           2e+04 &    8e+04 & 1.9e+02 & 9.4e+03 &   1.4e+03 & 8.8e+03 &  5.3e+02 \\
63  &                    LLOYDS BANKING GROUP PLC &           GB091 &              94 &  5.8e+04 &       0 &      32 &         0 &      62 &        0 \\
64  &                            DEUTSCHE BANK AG &           DE017 &         1.3e+04 &  5.5e+04 & 1.8e+03 & 7.7e+03 &   1.8e+02 & 2.6e+03 &  5.3e+02 \\
65  &                            SOCIETE GENERALE &           FR016 &         1.8e+04 &  5.1e+04 & 2.8e+03 & 8.8e+03 &     9e+02 & 4.8e+03 &  9.8e+02 \\
66  &                                        BPCE &           FR015 &         8.5e+03 &  4.1e+04 & 1.3e+03 & 5.4e+03 &   3.5e+02 &   1e+03 &  3.4e+02 \\
67  &        BBVA &           ES060 &         6.1e+04 &    4e+04 & 1.3e+02 & 4.2e+03 &   6.6e+02 & 5.6e+04 &        0 \\
68  &                      BANK OF VALLETTA (BOV) &           MT046 &              24 &     -&      10 &     3.9 &       2.8 &       0 &        7 \\
69  &                               BANCO BPI, SA &           PT056 &         5.5e+03 &  8.2e+02 & 3.2e+02 & 9.7e+02 &   3.9e+03 &       0 &  2.8e+02 \\
70  &                        BANCO SANTANDER S.A. &           ES059 &         5.1e+04 &  2.6e+04 & 1.8e+02 & 7.2e+02 &   3.7e+03 & 4.6e+04 &        0 \\
71  &                CAIXA DE AFORROS DE GALICIA, &           ES067 &         4.7e+03 &  2.3e+04 &  0.0022 & 1.6e+02 &   1.3e+02 & 4.4e+03 &        0 \\
72  &    CAIXA D'ESTALVIS DE CATALUNYA &           ES066 &         2.8e+03 &  2.3e+04 &       0 &       0 &         0 & 2.8e+03 &        0 \\
73  &                 CAJA DE AHORROS Y PENSIONES &           ES062 &         3.7e+04 &  2.2e+04 &       0 & 1.3e+03 &        26 & 3.5e+04 &        0 \\
74  &                                    KBC BANK &           BE005 &         7.9e+03 &  1.7e+04 & 4.4e+02 & 5.6e+03 &   1.6e+02 & 1.4e+03 &  2.7e+02 \\
75  &                      ERSTE BANK GROUP (EBG) &           AT001 &         1.2e+03 &  1.5e+04 & 3.5e+02 &   6e+02 &     1e+02 & 1.4e+02 &       40 \\
76  &                                   JP MORGAN &             JPM &              17 &  1.5e+05 &       0 &       0 &        17 &       0 &        0 \\
77  &                       BAYERISCHE LANDESBANK &           DE021 &         1.3e+03 &  1.4e+04 & 1.5e+02 & 5.1e+02 &   1.1e-05 & 6.6e+02 &       20 \\
78  &                                  BFA-BANKIA &           ES061 &         2.5e+04 &  1.2e+04 &      55 &       0 &         0 & 2.5e+04 &        0 \\
79  &                                 SNS BANK NV &           NL050 &           1e+03 &  5.4e+03 &      47 & 7.6e+02 &         0 &      57 &  1.6e+02 \\
80  &         RAIFFEISEN BANK (RBI) &           AT002 &         4.6e+02 &  1.1e+04 &     1.7 & 4.5e+02 &       2.1 &     3.5 &  0.00016 \\
81  &  DZ BANK AG DT. &           DE020 &         8.7e+03 &  1.1e+04 & 7.3e+02 & 2.7e+03 &     1e+03 & 4.2e+03 &       51 \\
82  &                              F VAN LANSCHOT &        Lanschot &              18 &  7.4e+02 &       0 &       0 &         0 &       0 &       18 \\
83  &                      ALLIED IRISH BANKS PLC &           IE037 &         6.5e+03 &  1.4e+04 &      40 & 8.2e+02 &   2.4e+02 & 3.3e+02 &    5e+03 \\
84  &     SKANDINAVISKA ENSKILDA BANKEN &           SE085 &         6.3e+02 &  1.2e+04 & 1.2e+02 & 2.9e+02 &   1.3e+02 &      86 &        0 \\
85  &                                    IBERCAJA &        Ibercaja &         9.6e+02 &  2.7e+03 &       0 &       0 &         0 & 9.6e+02 &        0 \\
86  &                LANDESBANK BADEN-WURT... &           DE019 &         2.8e+03 &  9.5e+03 & 7.8e+02 & 1.4e+03 &        95 & 5.4e+02 &        0 \\
87  &                 BANCO POPULAR ESPANOL, S.A. &           ES064 &         9.7e+03 &  9.1e+03 &       0 & 2.1e+02 &   6.4e+02 & 8.9e+03 &        0 \\
88  &       CAJA ESP. DE INVER. SALAMANCA &           ES070 &         7.6e+03 &     -&       0 &       0 &        27 & 7.6e+03 &        0 \\
89  &                NORDDEUTSCHE LANDESBANK &           DE022 &         2.8e+03 &  6.5e+03 & 1.5e+02 & 1.9e+03 &   2.6e+02 &   5e+02 &       41 \\
90  &                           BANCA MARCH, S.A. &           ES079 &         1.5e+02 &  6.5e+03 &       0 &       0 &         0 & 1.5e+02 &        0 \\
91  &                            OP-POHJOLA GROUP &           FI012 &              43 &  6.2e+03 &     3.1 &    0.36 &   0.00093 &    0.07 &       40 \\
92  &          BANCO COMERCIAL PORTUGUES, &           PT054 &         7.4e+03 &  4.4e+03 & 7.3e+02 &      50 &   6.5e+03 &       0 &  2.1e+02 \\
93  &                     BANCO DE SABADELL, S.A. &           ES065 &         7.4e+03 &  5.9e+03 &       0 &       0 &        91 & 7.3e+03 &       38 \\
94  &                HYPO REAL ESTATE HOLDING AG, &           DE023 &         1.1e+04 &     -&       0 & 7.1e+03 &   4.9e+02 & 3.4e+03 &       44 \\
95  &                       FRANKLIN ADVISERS INC &   Franklin Adv. &         3.6e+02 &  4.7e+02 &       0 &       0 &         0 &       0 &  3.6e+02 \\
96  &                            ABN AMRO BANK NV &           NL049 &         1.5e+03 &  2.8e+02 &       0 & 1.3e+03 &         0 & 1.1e+02 &  1.3e+02 \\
97  &                               MUENCHENER RV &       Munich RV &         8.2e+03 &  2.3e+04 & 5.8e+02 & 3.6e+03 &   4.2e+02 & 1.9e+03 &  1.8e+03 \\
98  &                    HSH NORDBANK AG, HAMBURG &           DE025 &           1e+03 &  4.8e+03 &   1e+02 & 6.6e+02 &        62 & 1.8e+02 &        0 \\
99  &                          GRUPO BANCA CIVICA &           ES071 &         4.8e+03 &     -&     5.4 &       0 &         0 & 4.7e+03 &        0 \\
100 &                CAIXA GERAL DE DEPOSITOS, SA &           PT053 &         6.8e+03 &  5.3e+03 &      51 &       0 &   6.5e+03 &   2e+02 &       23 \\
101 &            CAJA DE AHORROS DEL MEDITER... &           ES083 &         5.6e+03 &  3.8e+03 &       0 &      20 &       4.8 & 5.6e+03 &       15 \\
102 &                                   GRUPO BMN &           ES068 &         3.7e+03 &     -&       0 &       0 &        88 & 3.6e+03 &        0 \\
103 &                             BANK OF IRELAND &           IE038 &         5.6e+03 &    1e+04 &       0 &      30 &         0 &       0 &  5.6e+03 \\
104 &   DEKABANK &           DE028 &           6e+02 &  3.3e+03 &      87 & 2.7e+02 &        32 & 1.8e+02 &       30 \\
105 &                                       DEXIA &           BE004 &         2.3e+04 &  3.3e+03 & 3.5e+03 & 1.6e+04 &   1.9e+03 & 1.5e+03 &     0.34 \\
106 &                                   GRUPO BBK &           ES075 &         3.1e+03 &     -&       0 &       0 &         3 & 3.1e+03 &        4 \\
107 &                             BANKINTER, S.A. &           ES069 &         3.6e+03 &  3.1e+03 &       0 &     1.2 &         0 & 3.6e+03 &        0 \\
108 &                       WESTLB AG, DUSSELDORF &           DE024 &         2.2e+03 &    3e+03 & 3.4e+02 & 1.1e+03 &         0 & 7.5e+02 &       35 \\
109 &              UNIONE DI BANCHE ITALIANE SCPA &           IT044 &         1.1e+04 &  1.1e+04 &      25 & 1.1e+04 &         0 &       0 &        0 \\
110 &                      CAJA DE AHORROS Y M.P. &           ES072 &         3.3e+03 &  2.7e+03 &       0 & 3.8e+02 &         0 & 2.9e+03 &        0 \\
111 &             CAIXA D'ESTALVIS UNIO DE CAIXES &           ES076 &         2.6e+03 &     -&       0 &      11 &         0 & 2.6e+03 &       13 \\
112 &                    BANK OF CYPRUS PUBLIC CO &           CY007 &         2.8e+03 &  2.4e+03 & 2.4e+03 &      36 &         0 &      58 &  3.2e+02 \\
113 &                        LANDESBANK BERLIN AG &           DE027 &         1.1e+03 &  2.3e+03 & 4.5e+02 & 3.3e+02 &         0 & 3.7e+02 &    0.075 \\
114 &                                  ALPHA BANK &           GR032 &         5.5e+03 &    2e+03 & 5.5e+03 &       0 &         0 &       0 &        0 \\
115 &                             UNICREDIT S.P.A &           IT041 &         5.2e+04 &  9.3e+05 & 6.7e+02 & 4.9e+04 &        94 & 1.9e+03 &       58 \\
116 &               MARFIN POPULAR BANK PUBLIC CO &           CY006 &         3.4e+03 &  1.7e+03 & 3.4e+03 &       0 &         0 &       0 &       39 \\
117 &                          BANCO PASTOR, S.A. &           ES074 &         2.6e+03 &  1.6e+03 &      41 &   1e+02 &   1.2e+02 & 2.3e+03 &        0 \\
118 &                                 GRUPO CAJA3 &           ES078 &         1.5e+03 &     -&       0 &       0 &         0 & 1.5e+03 &      8.4 \\
119 &                   TT HELLENIC POSTBANK S.A. &           GR035 &         5.3e+03 &  9.3e+02 & 5.3e+03 &       0 &         0 &       0 &        0 \\
120 &                  EFG EUROBANK ERGASIAS S.A. &           GR030 &         8.9e+03 &  8.8e+02 & 8.8e+03 &   1e+02 &         0 &       0 &        0 \\
121 &          ESPIRITO SANTO GROUP, &           PT055 &         3.1e+03 &  6.2e+03 & 3.1e+02 &       0 &   2.7e+03 &      55 &        0 \\
122 &            AGRICULTURAL BANK OF GREECE &           GR034 &         7.9e+03 &  7.5e+02 & 7.9e+03 &       0 &         0 &       0 &        0 \\
123 &                  CAJA DE AHORROS DE VITORIA &           ES080 &           6e+02 &     -&       0 &       0 &         0 &   6e+02 &        0 \\
124 &                                 ING BANK NV &           NL047 &         1.1e+04 &  3.5e+04 & 7.5e+02 & 7.7e+03 &   7.6e+02 & 1.9e+03 &       92 \\
125 &                          RABOBANK NEDERLAND &           NL048 &         1.1e+03 &     -& 3.8e+02 & 4.4e+02 &        82 & 1.6e+02 &       60 \\
126 &                        WUERTTEMBERGISCHE LV &     Wuetter. LV &         7.7e+02 &  1.2e+02 &      85 & 4.5e+02 &        52 & 1.8e+02 &        8 \\
127 &                                    NYKREDIT &           DK011 &         1.1e+02 &     -&      22 &      88 &         0 &       0 &        0 \\
128 &                      MONTE DE PIEDAD Y CAJA &           ES073 &         3.3e+03 &       58 &       6 & 3.1e+02 &         0 & 2.9e+03 &        0 \\
129 &                      CAJA DE AHORROS Y M.P. &           ES081 &               6 &       58 &       0 &       0 &         0 &       6 &        0 \\
130 &                   BANCA MONTE DEI PASCHI DI &           IT042 &         3.3e+04 &  1.9e+03 &     8.1 & 3.2e+04 &     2e+02 & 2.8e+02 &        0 \\
131 &               COLONYA - CAIXA D'ESTALVIS DE &           ES082 &              26 &     -&       0 &       0 &         0 &      26 &        0 \\
132 &               BANQUE ET CAISSE D'EPARGNE DE &           LU045 &         2.8e+03 &  2.9e+03 &      85 & 2.4e+03 &   1.8e+02 & 1.7e+02 &        0 \\
133 &                          PIRAEUS BANK GROUP &           GR033 &         8.2e+03 & -1.9e+03 & 8.2e+03 &       0 &         0 &       0 &        0 \\
134 &                     NATIONAL BANK OF GREECE &           GR031 &         1.9e+04 & -4.3e+03 & 1.9e+04 &       0 &         0 &       0 &       18 \\
135 &                            ZURICH FINANCIAL &         Zurich  &         8.7e+03 &  2.5e+04 &       0 & 4.2e+03 &   3.7e+02 & 3.7e+03 &  3.7e+02 \\
136 &                                      MITSUI &          Mitsui &         6.4e+02 &  6.9e+04 &       0 & 3.7e+02 &        25 & 1.7e+02 &       76 \\
\end{longtable}
}


\end{document}